\documentclass[%
preprint,
amsmath,amssymb,
floatfix,
]{revtex4-1}
\usepackage{graphicx}
\usepackage{placeins}
\usepackage{dcolumn}
\usepackage{bm}
\usepackage[utf8]{inputenc}
\usepackage[T1]{fontenc}
\usepackage{mathptmx}
\usepackage{xcolor}
\usepackage{slashbox}

\begin{document}
\title[]{Vortex merging in strongly coupled dusty plasmas using a visco-elastic fluid model}
\author{Vikram Dharodi\footnote{Current affiliation: Department of Physics and Astronomy, West Virginia University, Morgantown, WV, USA}}
\email{vsd0005@auburn.edu}
\author{Evdokiya Kostadinova}
\email{egk003@auburn.edu}
\affiliation{Department of Physics, Auburn University, Auburn, Alabama 32849, USA}
	\date{\today}
\begin{abstract}
This work is a numerical study of the two-dimensional merging phenomena between  two Lamb-Oseen co-rotating vortices in a viscoelastic fluid. We use a generalized hydrodynamics fluid model to study vortex merging in a strongly coupled dusty plasma medium, which exhibits characteristics similar to a viscoelastic fluid.  Several aspects influencing the merging phenomena are considered: the aspect ratio (core size/separation distance), the relative circulation strengths of each vortex, and the coupling strength of the medium. Unlike classical hydrodynamic fluids, we find that for viscoelastic fluids, shear waves facilitate the merging events even for widely separated vortices. The merging process is accelerated in media with higher coupling strengths, but the resultant vortex shape decays more quickly as well. It is also found that varying either the vortex scale or the vortex circulation strength can result in a similar merging process, where a smaller (larger) vortex acts like a vortex with weaker (stronger) circulation. Finally, we show that a Poynting-like conservation theorem is satisfied for the examined merging processes. 
\end{abstract}
	\maketitle
\section{Introduction}\label{Introduction}
 \paragraph*{}
When a flowing fluid becomes non-laminar, it leads to the formation of different vortices \cite{dharodi2014visco,tur2017coherent},  such as rotating monopoles \cite{dharodi2020rotating}, propagating counter-rotating vortex pairs \cite{dharodi2016sub}, merging co-rotating vortex pairs \cite{leweke2016dynamics}, and tripoles \cite{rossi1997quasi,kizner2004tripole}. A medium associated with such different types of vortices can lead to a two-dimensional turbulent flow, where the merging of small-scale vortices generates larger ones, resulting in energy transfer to larger scales (i.e., a reverse cascade) \cite{couder1989hydrodynamics,mcwilliams1990vortices,vincent1991spatial,hopfinger1993vortices,jimenez1996structure}.  Vortex merging is ubiquitous in hydrodynamic fluids \cite{overman1982evolution}, geophysical flows \cite{gerosa2017merging,reinaud2018merger}, plasma flows \cite{kendl2018gyrofluid}, astrophysical systems \cite{ebisuzaki1991merging}, and aeronautics \cite{chen1999dynamics,jacquin2005unsteadiness}. Numerous studies have been conducted on the merging phenomena in basic hydrodynamic (HD) flows, including the merger of two symmetric vortices \cite{leweke2001controlled,dritschel2002vortex,meunier2002merging,josserand2007merging}, asymmetric vortices \cite{jing2012insights,chen2020experiments,folz2023asymmetric}, and unequal size vortices \cite{dritschel1992quantification,trieling2005interaction}. The main objective of this article is to understand how the merging process occurs in viscoelastic (VE) fluids where viscosity contributes to the transverse mode in the presence of elasticity \cite{frenkel_kinetic}.
 \paragraph*{}
Of specific interest to this study is a strongly coupled dusty plasma (SCDP) which is an ideal VE medium for laboratory investigations of rotating vortices and their interactions \cite{agarwal2003spontaneous,vaulina2004analysis,vaulina2004formation,feng2007observation,chai2016vortex,choudhary2017experimental,choudhary2018collective,hartmann2019self,melzer2021physics,kumar2023kelvin,ramkorun2023introducing}. A complex (or dusty) plasma is a collection of microparticles suspended in low-temperature, weakly ionized gas. The dust particles collect ions and electrons from the plasma and can exhibit self-organization into strogly-coupled structures. Below the crystallization limit, a SCDP behaves like a VE fluid \cite{ikezi1986coulomb,vladimirov1997vibrational}. The elastic behavior in SCDP is induced by the strong correlations among the dust grains, due to which this medium supports the incompressible transverse shear modes in addition to the compressible longitudinal modes \cite{Kaw_Sen_1998}. Additionally, the presence of elasticity diminishes the viscosity's pure damping effect, allowing a vortex structure to survive in dusty plasmas longer than in a simple HD fluid (where viscosity exhibits a pure damping effect). 

Molecular dynamics simulations of SCDP monolayers have demonstrated the onset of turbulence and reverse energy cascade in laboratory-relevant conditions \cite{schwabe2017turbulence,thomas2018complex}. Experiments from the Plasmakristall-3 (PK-3) facility on board of the ISS have demonstrated that the onset of the heartbeat instability and turbulence changes the dust particle kinetic energy spectrum from an exponential to a power-law, suggesting a double energy cascade predicted for 2D forced turbulence \cite{mccabe2022investigation}. Experiments from the Plasmakristall-4 (PK-4) facility, the PK-3 successor on the ISS, have revealed an unusual redistribution of dust kinetic energy at the onser of polarity switching of the DC current \cite{kostadinova2021fractional}. All these examples suggest that energy transfer in SCDPs, and other VE fluids, differ substantially from what is expected for ordinary HD fluids. A detailed investigation of the energy distribution during vortex merging in SCDPs can provide crucial information on the physical mechanisms causing the energy cascades observed in these systems.
 \paragraph*{} 
This research uses a well-known phenomenological generalized hydrodynamics (GHD) fluid model to simulate the SCDPs as VE fluid media \cite{Kaw_Sen_1998, Kaw_2001,tiwari2014evolution,tiwari2015turbulence,gupta2018compressibility}. In this model, the VE effects are characterized by two coupling parameters: the Maxwell relaxation value $\tau_m$ and the shear viscosity $\eta$. The ratio of $\eta/{\tau_m}$ is proportional to the coupling strength of the VE fluid, which determines the viscoelastic nature of the medium. To examine the exclusive impact of transverse modes on merging events and to make sure they don't obstruct longitudinal modes, we consider the incompressible limit of the GHD model (i-GHD). Transverse modes in the dusty plasma medium have been investigated analytically \cite{peeters1987wigner,vladimirov1997vibrational,wang2001longitudinal}, experimentally \cite{nunomura2000transverse, pramanik2002experimental} and computationally \cite{schmidt1997longitudinal}. Using the GHD model, the presence of shear waves has been theoretically predicted by Kaw et al. \cite{Kaw_Sen_1998,Kaw_2001}. Applying the incompressible limit to the GHD model, Dharodi et al. \cite{dharodi2014visco} have computationally predicted the shear waves and their role in various physical processes. Those include their ability to suppress instabilities brought on by gravity (Rayleigh-Taylor and buoyancy-driven instability)~\cite{das2014collective,dharodi2021numericalA,dharodi2021numericalB}, to mix VE fluids more effectively than standard HD fluids in rotating sharp shear flows \cite{tiwari2014kelvin,dharodi2022kelvin}, to control the evolution of density spiral waves in heterogeneous density fluids~\cite{dharodi2020rotating}, and to influence the propagation of dipole structures~\cite{dharodi2016sub}. The occurrence of shear waves on the charged VE fluid surface can have important role in the electrostatic breakdown of droplets in misty plasma~\cite{coppins2010electrostatic}, which has important application to plasma-enhanced chemical vapour deposition and fusion plasmas.

 \paragraph*{}
 This paper explores various factors that influence the merging phenomenon between two Lamb-Oseen co-rotating vortices, including the coupling strength ($\eta/{\tau_m}$) of the medium, the relative circulation strengths of each vortex, and the aspect ratio (core size/separation distance). In simple HD fluids, the merging phenomenon is expected to occur when the aspect ratio is greater than the merger threshold value. We study the two cases of initial aspect ratio: one where the ratio is less than the merger threshold (widely spaced) and another where the threshold is exceeded (closely spaced). For each separation case, we have simulated three different circulation strengths: strong, medium, and weak.  For each circulation, we have considered three coupling strengths: mild-strong, medium-strong, and strongest. This makes a total of 18 sets of initial conditions for two vortices of the same size.   
 \paragraph*{}
 For the widely spaced case, for strong circulation, we have observed merging phenomena for all three coupling strengths. For medium circulation, the merging phenomenon is observed for mild-strong coupling only while in medium-strong and strongest coupling strengths, the vortex-pair disappears due to the emergence of shear waves that remove energy from each vortex at a faster rate. For weak circulation, the vortex-pair disappears for all three coupling strengths prior to the merger event. It is observed that the interaction between a vortex pair (merging or disappearance) is governed by a competition between the coupling strength of the medium and the circulation strength of the vortices. For the closely spaced case, for medium circulation, the merging phenomenon is observed for mild-strong and medium-strong coupling strengths, but the vortex pair disappears at the strongest coupling strengths. When two vortices have different circulation strengths, the weaker vortex deforms faster than the stronger vortex. The merging process of two vortices with different relative sizes is strikingly similar to that of vortices with different strengths, where the larger vortex behaves like a vortex with strong circulation and the smaller vortex acts like a vortex with weaker circulation. 
 \paragraph*{}
 We re-visit a Poynting-like conservation theorem drived in our previous work [Phys. Plasmas 23, 013707 (2016)], and employ it in the merging processes. Our numerical results demonstrate that this theorem is precisely satisfied by quantifying the radiative, convective, and dissipative quanties that cause changes in the conserved quantity W. 
 \paragraph*{}
The structure of this paper is as follows:  Sec. \ref{ghd_model} describes the i-GHD model and its normalizing process. In addition, we talk about the simulations methodology (Sub Sec. \ref{methodology}) and  we re-drive a conserved quantity and an equation (Sub Sec. \ref{Conserved_quantity}). The numerical results are discussed in Sec. \ref{Results_discussion}. Finally, we analyze our findings and provide closing thoughts in Sec. \ref{Conclusions}.
\section{Incompressible generalized hydrodynamics (i-GHD) fluid model} \label{ghd_model}
 \paragraph*{}
A strongly coupled dusty plasma exhibits VE fluid behavior below the crystallization limit. The generalized hydrodynamic fluid model is used in this study to examine this SCDP~\cite{Kaw_Sen_1998, Kaw_2001,tiwari2014evolution,dharodi2014visco,tiwari2015turbulence,dharodi2016sub}. The coupling strength of this medium fluid is proportional to the ${\eta}/{\tau_m}$ ratio~\cite{frenkel_kinetic}. Viscosity $\eta$ in the presence of elasticity $\tau_m$ (relaxation time parameter) contributes to the transverse mode. Both compressible longitudinal and incompressible transverse shear modes are supported by this model. We completely isolate the compressibility effects in order to focus on the incompressible aspects of this system. In the incompressible limit (i-GHD), the normalized continuity and momentum equations for the dust fluid  can be written as:
\begin{equation}\label{eq:continuity}
  \frac{\partial \rho_d}{\partial t} + \nabla \cdot
\left(\rho_d\vec{v}_d\right)=0{,}
  \end{equation}
\begin{eqnarray}\label{eq:momentum1}
	&&\left[1+{\tau_m}\left(\frac{\partial}{\partial{t}}+{\vec{v}_d}\cdot \nabla\right)\right]\nonumber\\
	&& \left[ {{\rho_d}\left(\frac{\partial{\vec{v}_d}}{\partial {t}}+{\vec{v}_d}{\cdot} \nabla{\vec{v}_d}\right)}+{\nabla}{p_d}+{\rho_c}\nabla \phi_{d} \right]\nonumber\\
	&& =\eta \nabla^2\vec{v}_d{,}
\end{eqnarray}
correspondingly, and the incompressible state is stated as 
\begin{equation}
\label{eq:incompressible}
{\nabla}{\cdot}{\vec{v}_d}=0{.}
\end{equation}
In our previous articles \cite{dharodi2014visco,dharodi2016sub}, we have covered in depth the derivation of these normalized equations as well as the process of their numerical implementation and validation. We take into account a constant charge density, which could be negative or positive. Although dust particles are normally negatively charged, they can become positively charged under certain situations, such as secondary electron emission \cite{shukla2015introduction} or in afterglow plasma conditions \cite{chaubey2021positive,chaubey2022coulomb,chaubey2022preservation}. In the afterglow, based on the charge of dust grains, one can control the fall of dust particles \cite{chaubey2023controlling,chaubey2023mitigating}. Since the charge density fluctuations are ignored in the incompressible limit, the Poisson equation is replaced by the quasi-neutrality condition. In this case, the mass density of the dust fluid is ${\rho_d}= {n_d}{m_d}$, the number density of the dust fluid is $n_d$, and the mass of the dust particle is $m_d$. The dust charge density, dust fluid velocity, and dust charge potential are represented by the variables $\rho_c$, $\vec{v}_d$, and $\phi_d$, respectively. The electric field is the primary cause of the pressure $p_d$ term in a strongly coupled plasma \cite{flanagan2010observation}. The inverses of the dust plasma frequency $\omega^{-1}_{pd} = \left({4\pi (Z_d e)^{2}n_{d0}}/{m_{d0}}\right)^{-1/2}$, plasma Debye length $\lambda_{d} = \left({K_B T_i}/{4{\pi} {Z_d}{n_{d0}}{e^2}}\right)^{1/2}$, ${\lambda_d}{\omega_{pd}}$, and ${{Z_d}e}/{{K_B}{T_i}}$, respectively, are used to normalize the time, length, velocity, and potential. The dust grain mass, ion temperature, and Boltzmann constant are denoted by the parameters $m_d$, $T_i$, and $K_B$, respectively. Without taking charge fluctuation into account, the charge on each dust grain, $Z_d$. The equilibrium value $n_{d0}$ normalizes the number density $n_d$.
In the case of constant density $i.e$~ $  {\rho_d}(x,y,t)={\rho_{cd}}$, equation \ref{eq:momentum1} becomes
\begin{equation}\label{eq:momentum2}
 {\left[1 + \tau_m \left(\frac{\partial}{\partial t}+\vec{v}_d  \cdot \nabla
\right)\right]  \left[{\frac{\partial \vec{v}_d } {\partial t}+\vec{v}_d  \cdot{\nabla}\vec{v}_d } + \frac{\nabla {p_d}}{\rho_{cd}} -\frac{\rho_c}{\rho_{cd}} {\nabla {\phi_d}} \right]  =  {\eta'}\nabla^2 \vec{v}_d}{.}
  \end{equation}
  where ${\eta'}={\frac{\eta}{\rho_{d}}}$. In limit {${\tau_m}{\frac{\partial}{\partial{t}}} \geq 1$}, if we take the curl of above momentum Eq.~(\ref{eq:momentum2}) and keep only the linearized terms, we found that the i-GHD model supports transverse waves moving with phase velocity ${v_p}=\sqrt{{\eta}/{{\rho_d}{\tau_m}}}$. 
\subsection{Simulation methodology}
\label{methodology}
\paragraph*{}
In order to carry out numerical simulation the generalized momentum equation~(\ref{eq:momentum2}) has been formulated as a set of two coupled convective equations 
\begin{eqnarray}\label{eq:vort_incomp1}
 {\frac{\partial \vec{v}_d } {\partial t}+\vec{v}_d  \cdot{\nabla}\vec{v}_d } + \frac{\nabla {p_d}}{\rho_{cd}} -\frac{\rho_c}{\rho_{cd}} {\nabla {\phi_d}} ={\vec \psi}
\end{eqnarray}
\begin{equation}\label{eq:psi_incomp1}
	\frac{\partial {\vec \psi}} {\partial t}+\vec{v}_d \cdot \nabla{\vec \psi}=
	{\frac{\eta'}{\tau_m}}{\nabla^2}{\vec{v}_d }-{\frac{\vec \psi}{\tau_m}}{.}
\end{equation}
It is evident from Eq.~(\ref{eq:vort_incomp1}) that the strain produced in the elastic medium by the time-varying velocity fields is represented by the quantity ${\vec \psi}(x,y)$. Take the curl of equation~(\ref{eq:vort_incomp1}). The curls of third and forth terms vanish because, because we have assumed a constant number and a constant charge density, a gradient's curl is zero. Equation (\ref{eq:vort_incomp1}) is transformed into:
  \begin{eqnarray}\label{eq:vort_incomp3} 
\frac{\partial{\vec \omega}} {\partial t}+\vec{v}_d \cdot \nabla{{\vec \omega}}
={\nabla}{\times}{\vec \psi}{,}
\end{eqnarray}
\begin{eqnarray}\label{eq:psi_incomp3}
\frac{\partial {\vec \psi}} {\partial t}+\vec{v}_d \cdot \nabla{\vec \psi}=
{\frac{\eta'}{\tau_m}}{\nabla^2}{\vec{v}_d }-{\frac{\vec \psi}{\tau_m}}{.}
\end{eqnarray}
The set of momentum equations (\ref{eq:vort_incomp3}) and (\ref{eq:psi_incomp3}) have been solved numerically using the LCPFCT approach (Boris {\it et al.}~\cite{boris_book}). This approach is predicated on a finite difference scheme connected to the flux-corrected algorithm. In order to satisfy the incompressible criterion, Poisson's equation $ {\nabla^2}{\vec{v}_d}=-{\nabla}{\times}{\vec {\omega}}$ has been solved to update the corresponding velocity at each unit time step using the software FISPACK~\cite{swarztrauber1999fishpack}. During the whole simulation effort, boundary conditions are periodic in both the x- and y-direction. To ensure the grid independence of the numerical results, a grid convergence analysis has been conducted in each case.
 \subsection{Conserved quantity and equation}
 \label{Conserved_quantity}
\paragraph*{}
In order to drive the conserved equation, take dot products (\ref{eq:vort_incomp3}) by ${\eta'}/{\tau_m}{\times}\vec{\omega}$ and (\ref{eq:psi_incomp3}) by $\vec{\psi}$, then add them. Next, we take the integral of the resultant equation, which gives us
\begin{multline}\label{eq:integral_equ} 
\underbrace{{\frac{\partial}{\partial t}}
{\int_{V}}{\left(\frac{\psi^2}{2}+{\frac{\eta}{\tau_m}}\frac{\omega_z^2}{2}
\right)}dv}_{\text{\bf {dWdt}}}= \\
-\underbrace{{\frac{\eta}{\tau_m}}{\oint_{S}}({\omega_{z}}{{\times}{ \vec
\psi}}){\cdot}d{\bf{a}}}_{\text{
\bf S}}-\underbrace{{\oint_{S}}\left(\frac{\psi^2 }{2 }+{\frac{
\eta}{\tau_m}}\frac{\omega_z^2}{2}\right){\vec{v}_d }{\cdot}d{\bf{a}}}_{\text{\bf
T}}-\underbrace{{\int_{V}}{\frac{\psi^2}{\tau_m }}dv}_{\text{\bf P}}{.} 
 \end{multline}
  Here, ${\bf{dWdt}}$ is the sum of square integrals of vorticity ($\omega_z$) and the velocity strain ($\psi$, created in the elastic medium by the time varying velocity fields).  The radiation  term  ${\bf{S}}$ is like a Poynting flux for the radiation corresponding to the transverse shear waves. It accommodates the integral of the cross product of $\omega_z \hat{z}$ and $\vec{\psi}$. The other mechanism causing the change in $W$ is through convection ${\bf{T}}$ and the viscous dissipation ${\bf{P}}$ through $\eta$. 
\paragraph*{}
Note:  The appendix \ref{Conserved_quantity_A} has a complete, step-by-step derivation of this theorem. Our earlier paper \cite{dharodi2016sub}, along with the derivation, has numerical simulations that validate it for the single rotating vortex and dipole evolution of two unlike-sign vortices.

\section{Results discussion and numerical simulation}
\label{Results_discussion}
\paragraph*{}
The merging of of two like-signed vortices is a well-known phenomenon in a HD system. Rather than giving simulation findings based only on contour plot interpretations, let us first discuss the the different stages involved in merging phenomena in an HD system~\cite{josserand2007merging,trieling2010dynamics}. In an HD system, the stages that occur when two vortices are separated are as follows: both rotating vortices revolve around the vortex centroid as a whole. The local strain induced by each vortex on its companion causes the elliptical distortion of vortex outlines. Viscosity-based or numerical diffusion augments the core of both vortices. This effect modifies the aspect ratio of the separation distance to the vortex radius, leading to a critical point where merging takes place. When the critical  merger ratio is attained the merging happens quickly along with two spiral filaments ejected in an outer region of vortices.
\paragraph*{}

\subsection*{Single rotating vortex}
\label{monopole_evolution}
\paragraph*{}
A single circular Lamb–Oseen vortex is a solution for the Navier–Stokes equations. For a specific case of the Lamb–Oseen vortex the velocity profile is given as
\begin{equation}
\label{eq:dipole_profile} 
{U_{0}(r,t=0)}={\frac{\Gamma_0}{2\pi{r}}}{{\left(1-e^{-{r^2}/{a^2_{0}}}\right)}}
\end{equation}
 Here, $r^2={(x-x_0)^2+(y-y_0)^2}$. $a_0$ is the initial vortex radius and ${\Gamma_0}$ is the circulation of the vortex. In accordance with the velocity profile above, the vorticity is
\begin{equation}
\label{eq:dipole_profile} 
{\omega_{0}(r,t=0)}={\frac{\Gamma_0}{2\pi{a^2_0}}}{{e^{-{r^2}/{a^2_{0}}}}}{.}
\end{equation}
This smooth Gaussian rotating vortex drives the shear in flow. The smoothness of this vortex helps to prevent strong shear flows, which could cause Kelvin-Helmholtz instability~\cite{dharodi2022kelvin}. In our previous work, we numerically examined the dynamics of a typical Lamb-Oseen vortex (Figure 1 in \cite{dharodi2014visco}). We observed that the rotating vortex stays stable in the inviscid HD fluid, while in the VE fluid, the shear flow induces cylindrical TS waves that propagate radially outward at phase velocity $\sqrt{{\eta}/{\tau_m}}$ into the surrounding media, with amplitudes decreasing to $1/{\sqrt{r}}$.
 
\subsection*{Two co-rotating vortices}
\label{co-rotating_vortices}
\paragraph*{}
Our primary objective is to understand the phenomenon of merging, which occurs when two co-rotating Lamb-Oseen vortices are brought together. Hence, we consider a superposition of two co-rotating circular Lamb-Oseen vortices initializes the flow. So, the initial vorticity field becomes
 \begin{eqnarray}
 \label{eq:dipole_profile} 
 {\omega_{0}(r_1,r_2,t=0)}={\frac{\Gamma_1}{\pi{a^2_1}}}{{e^{-{r_1^2}/{a^2_{1}}}}}+
  {\frac{\Gamma_2}{\pi{a^2_2}}}{{e^{-{r_2^2}/{a^2_{2}}}}}{.}
 \end{eqnarray}
Here $r_1^2={\left(x-x_{01}\right)^2-\left(y-y_{01}\right)^2}$ and $r_2^2={\left(x-x_{02}\right)^2-\left(y+y_{02}\right)^2}$. $a_1$ and $a_2$ are the initial radii of two vorticies, $b_0={x_{02}}-{x_{01}}$ is the separation distance between the vortex cores, and the circulations of the two vorticies are $\Gamma_1$ and $\Gamma_2$. It should be pointed out that the interaction between vortices, in addition to the medium's nature (coupling strength; $\eta/{\tau_m}$), is governed by their strengths, sizes, and initial spacing. Next, we are going to discuss how these factors affect the merging phenomenon in our considered VE medium numerically. Table~\ref{table:table1} offers an overview of the cases that will be explored in the next sections.
\begin{table}[h]
 \begin{tabular}{|l||*{3}{c|}}\hline
\backslashbox{case no.}{{\hspace{-0.8cm}}case}
&\makebox[5em]{strength ($\Gamma$)}
&\makebox[5em]{size ($a$)}
&\makebox[5em]{spacing ($b_0$)}
\\\hline\hline
{A}& {equal}& {equal}& {widely and closely}\\\hline
{B}& {unequal}& {equal}& {widely}\\\hline
{C}& {equal} & {unequal}& {widely}\\\hline
\hline
\end{tabular}
\caption{All the three cases have the following varying coupling strengths:: the mild-strong ($\eta$=2.5, $\tau_m$=20), medium-strong ($\eta$=2.5, $\tau_m$=10), and strong or strongest ($\eta$=2.5, $\tau_m$=5). Additionally, the inviscid limit of HD fluids is shown for each scenario.}
\label{table:table1}
\end{table}
For each simulation that follows, the system of length $lx=ly=24{\pi}$ units contains $512{\times}512$ grid points on both the $x$ and $y$ axes. The system goes from $-12{\pi}$ to $12{\pi}$ units along the x- and y-axes. The vortices are placed horizontally: $({y_{01}}, {y_{02}})=(0,0)$. The colorbar in every vorticity contour plot indicates the strength of the vorticity. For each scenario, the inviscid limit of HD fluids is also demonstrated in order to provide a deeper understanding of the dynamics from a physical perspective.
\subsection{Equal strength and size: $\Gamma_1{=}\Gamma_2$ and ${a_1=a_2}$}
\label{eq_strength_eq_size}
 \paragraph*{}
  Here, we consider a pair of symmetric vortices of identical circulation, $\Gamma_1=\Gamma_2=\Gamma_0$, and equal initial vortex radius, ${a_1=a_2=a_0=\pi/2}$. For simple hydrodynamic fluids, the symmetric vortex merger has been extensively studied based on the initial aspect ratio $a_0/b_0$ between the vortex core radius $(a_0)$ and the vortex separation $(b_0)$ ~ \cite{griffiths1987coalescing,hopfinger1993vortices,leweke2001controlled,meunier2002merging,cerretelli2003physical}.  It is observed that a critical the critical merger ratio $({a_0}/{b_0})_{crit}=0.29$ results the vortex merger~ \cite{griffiths1987coalescing,hopfinger1993vortices,leweke2001controlled,meunier2002merging,cerretelli2003physical}. We examine the two initial aspect ratios: in arrangement A, the initial aspect ratio (0.167) is less than the critical ratio, and in arrangement B, the initial aspect ratio (0.33) is greater than the critical ratio.
  \subsubsection*{a. Widely spaced {$\left(a_0/b_0~{<}~({a_0}/{b_0})_{crit}\right)$}}
\label{eq_strength_eq_size_widely}
\begin{figure}[!ht]
 \centering               
\includegraphics[width=\textwidth]        {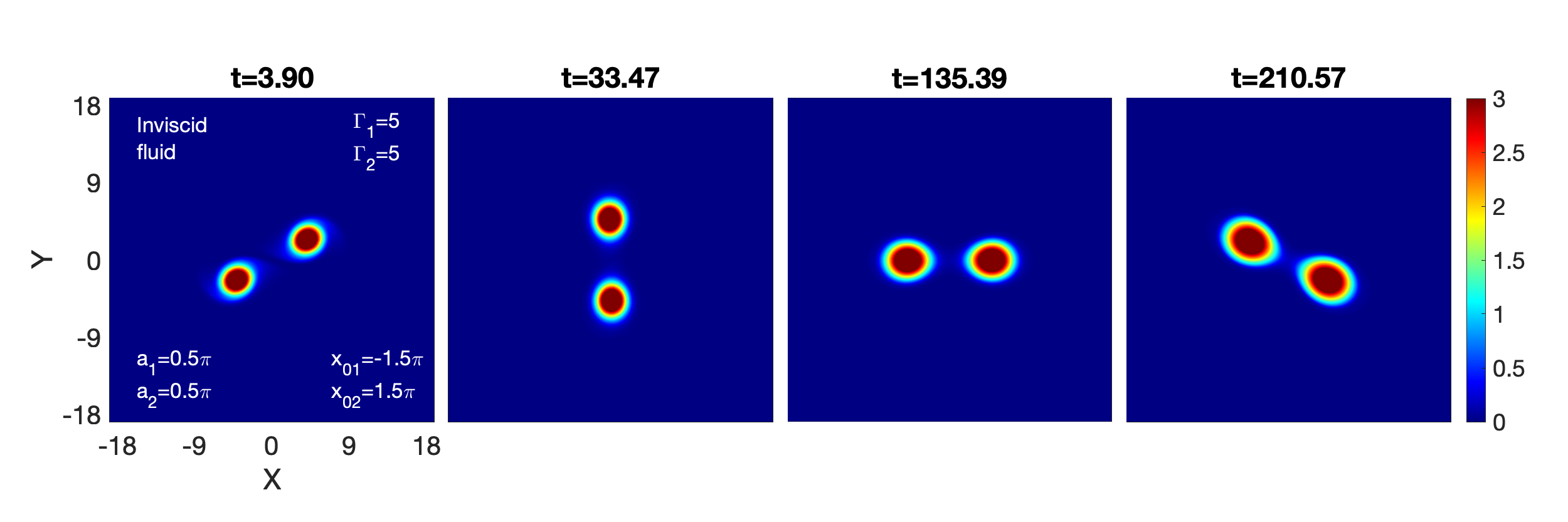}
        \caption{Time evolution of  vorticity for an inviscid HD fluid for $\Gamma=5$. The vortices rotate about each other due to the mutually induced velocity, no merging phenomena.}
\label{fig:figure1}
\end{figure}

 \paragraph*{}
Here, we consider $b_0=6a_0$ with $({x_{01}},{x_{02}})=(-3a_0,3a_0)$. This results in an aspect ratio of $0.167$, which is initially below the merger threshold of 0.29, indicating that the merging phenomena is not expected to occur. Three circulation strengths, namely: strong circulation $\Gamma_0=10$, medium circulation $\Gamma_0=5$, and weak circulation $\Gamma_0=3$—are simulated as cases (1), (2), and (3), respectively. 
 \paragraph*{}
First, for all three circulation strengths (10, 5, and 3), we have run the simulations for the incompressible inviscid fluids. Due to mutually induced velocity, we find that the two vortices spin around the total vortex centroid at varying speeds, according to their respective circulation strengths: rapid for $\Gamma_0=10$, sluggish for $\Gamma_0=5$, and slowest for $\Gamma_0=3$. Regardless of the circulation strengths, no evidence of merging events is found. It is evident from figure \ref{fig:figure1} that displays the progression of vorticity in the inviscid HD fluid for $\Gamma_0=5$. 
Subsequently, the merging process in SCDPs represented as VE fluids would be an intriguing observation. The SCDPs facilitate the radial emission of TS waves  from a rotating vorticity structure into the surrounding fluid. The TS waves have the same symmetry like the structure that generates them \cite{dharodi2014visco}.  The velocity ($v_p=\sqrt{\eta/\tau_m}$) of these shear waves is proportional to the medium's coupling strength $(\eta/{\tau_m})$. Stated differently, a medium possessing a higher coupling strength would be able to sustain the quicker TS waves. In SCDPs, for above three cases, the coupling strength is the only variable that varies with the fixed viscosity $\eta= 2.5$ through the elastic term $\tau_m$. The mild-strong ($\eta$=2.5, $\tau_m$=20), medium-strong ($\eta$=2.5, $\tau_m$=10), and strong or strongest ($\eta$=2.5, $\tau_m$=5) coupling strengths have been proposed in each case.  
\subsubsection{Strong circulation: $\Gamma_1=10$, $\Gamma_2=10$; $a_1=0.5\pi$, $a_2=0.5\pi$; $and$ $b_0={3\pi}$}
\label{eq_strength10_eq_size_widely}
\begin{figure}[!ht]
\centering               
\includegraphics[width=\textwidth]{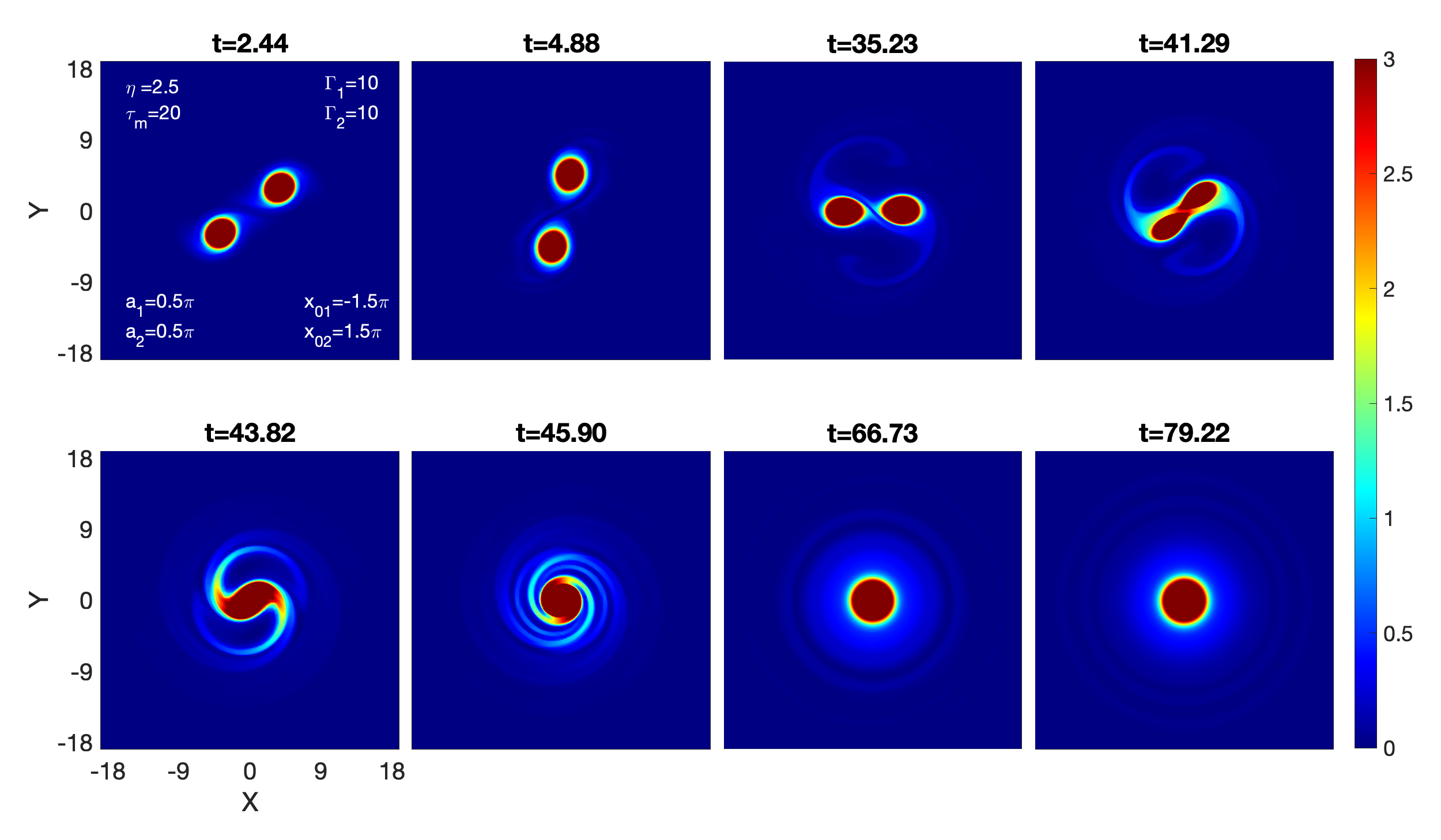}
\caption{Vorticity structures in the time of the merging event of two co-rotating vortices in a VE fluid. $\eta=2.5$; $\tau_m=20$ in this case. Both vortices have the same circulation strength, $\Gamma_0=10$, and are identical.}
\label{fig:figure2}
\end{figure}

\begin{figure}[!ht]
\centering               
\includegraphics[width=\textwidth]{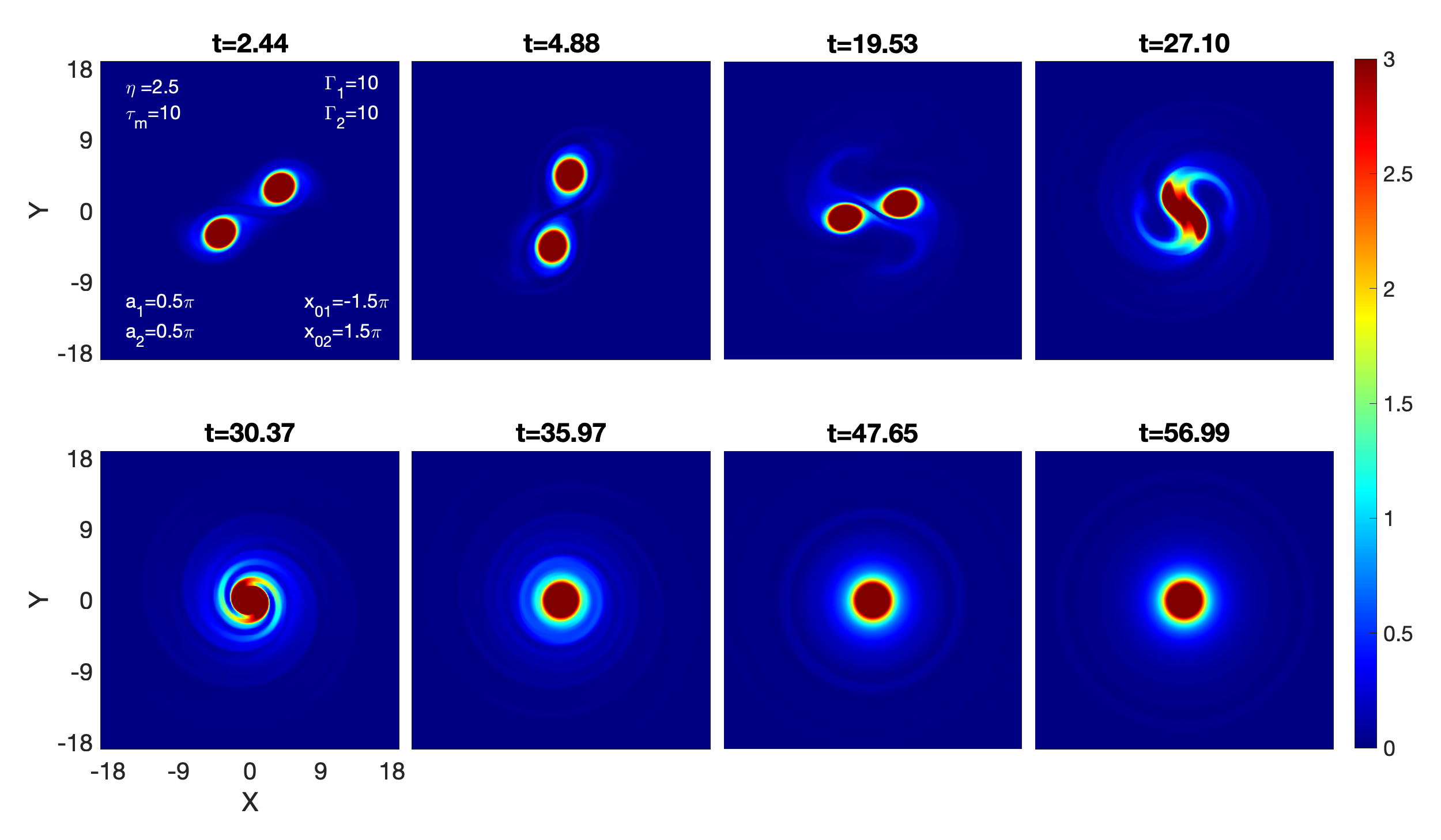}
\caption{The time evolution of an identical vortex pair, equal circulation strength $\Gamma_0=10$, in viscoelastic fluid ($\eta=2.5$; $\tau_m=10$).}
    \label{fig:figure3}
\end{figure}

\begin{figure}[!ht]
\centering               
\includegraphics[width=\textwidth]
{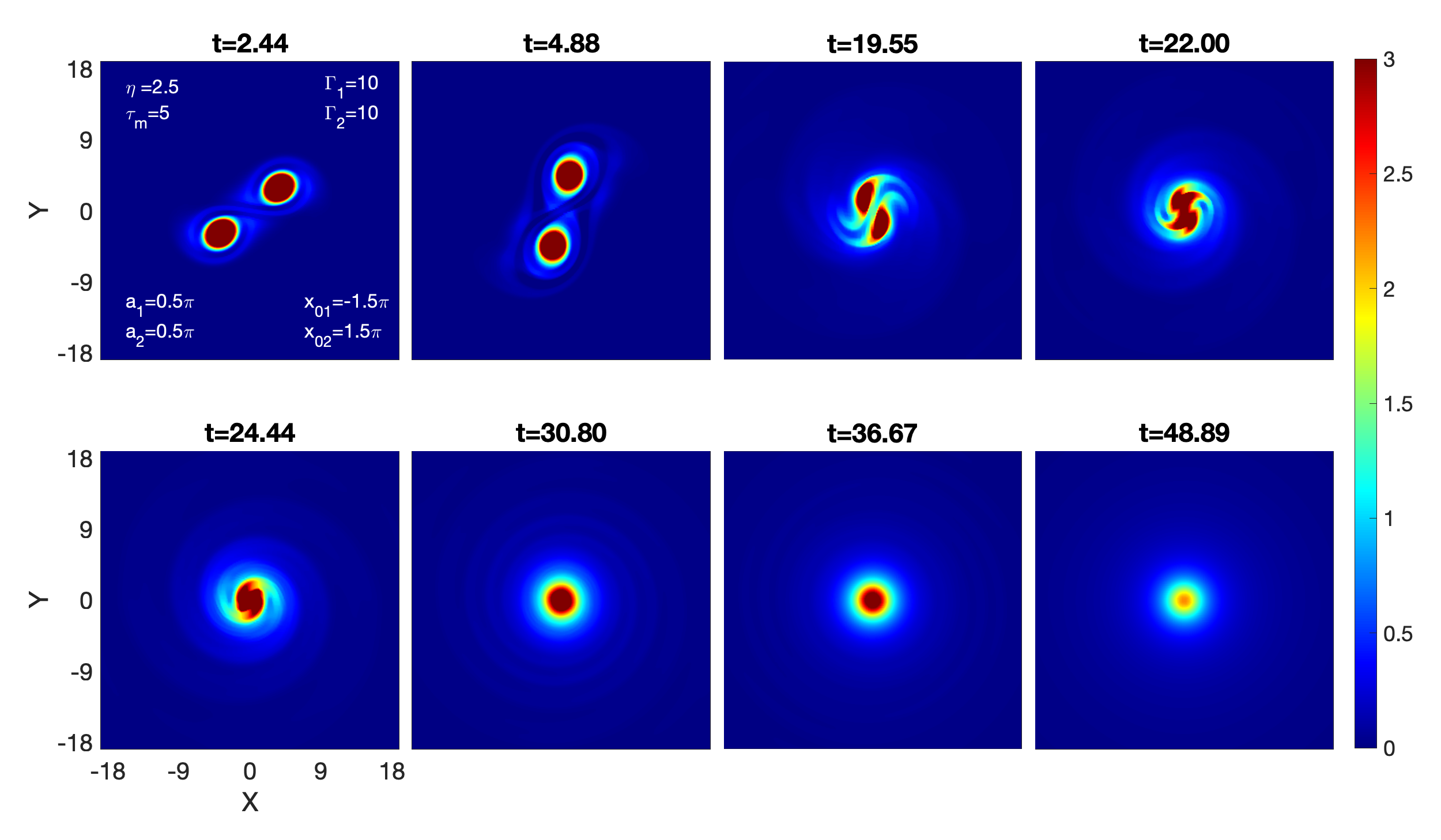}
\caption{The time evolution of an identical vortex pair, equal circulation strength $\Gamma_0=10$, in viscoelastic fluid ($\eta=2.5$; $\tau_m=5$).}
       \label{fig:figure4}
\end{figure}

\paragraph*{}
The vorticity field during the merging of two co-rotating vortices in a VE fluid, with coupling parameter values of $\eta=2.5$ and $\tau_m=20$, is depicted in Figure \ref{fig:figure2}. Unlike HD case (see figure~\ref{fig:figure1}), the spining co-rotating vortices merge into a single vortex. We know that, in addition to the rotational motion, this VE fluid favors the emission of shear waves with the phase velocity $v_p=\sqrt{\eta/\tau_m}$ = 0.35 into the surrounding fluid from both vortices in a radial pattern. The first panel at $t=2.44$ clearly illustrates the shear waves originating from the two vortices. It should be noted that the wavefronts from one vortex considerably impedes the wavefronts from the other, causing the wavefronts profile to become asymmetrically concentrated around each vortex. These wavefronts like wakes stay behind the counterclockwise rotating vortices and take the form of an "S" structure around the vortices which is visible in the second panel at $t=4.88$. As time progresses (from $t=35.23$ to  $t=79.22$), the continuous emission of waves, may be along with the some dragging viscous contribution, begin to deform into filaments, and the vortices start moving closer together over time. First, the vortex pair forms a non-symmetric  single vortex, encircled by spiral wavefronts, and eventually, this non-symmetric vortex transforms into a symmetric Gaussian vortex and the spiral wavefronts shape into cylindrical circular forms. The last two panels at $t=66.73$ and $t=79.22$ show the outgoing cylindrical TS waves from symmetric vortex, which is obviously distinct from pure viscous damping.
In conclusion, as the TS emerges outward, vortices get bigger. This, in turn, enhances the attraction between the vortices over time and modifies the aspect ratio. As the critical merger ratio is achieved, the vortices eventually merge into a single vortex.
\paragraph*{}
A novel VE fluid condition with $\eta = 2.5$ and $\tau_m = 10$ is simulated in figure \ref{fig:figure3}. This condition supports the shear waves moving at a phase velocity $v_p =\sqrt{\eta/\tau_m}$ = 0.5, which is greater than the above simulated scenario ($v_p =\sqrt{\eta/\tau_m}$ = 0.125; see figure \ref{fig:figure2}). A comparative analysis between the figure \ref{fig:figure2} and  figure \ref{fig:figure3} reveals that the merging process is nearly same, albeit happening more quickly in figure \ref{fig:figure3} than figure \ref{fig:figure2}. This acceleration of the merger process can only be explained by faster moving TS waves. It can be visualized from the comparative analysis of the panel at $t=47.65$ from figure \ref{fig:figure3} and the panel at $t=66.73$ from figure \ref{fig:figure2}.
A new VE fluid with a stronger coupling strength $(\eta = 2.5, \tau_m = 5.0)$ than the earlier scenarios is shown in figure \ref{fig:figure4}. This medium supports the shear waves at a phase velocity $v_p =\sqrt{\eta/\tau_m}$ = 0.71, which is faster than above discussed scenarios. Apart from the acceleration in the merging process, there are a few little variations, but overall, the merging process is nearly identical to that observed in earlier scenarios. Since, here, the shear waves are faster than the previous two scenarios, the waves originating from the two vorticies form an "eight" configuration surrounding them, as seen in the first panel at $t=2.44$. Subsequently, this wavefront expands and assumes the shape of an envelope encircling the entire pair of vortices (outer), while a new wavefront forms a new "eight" structure around the vortices (inner), are visible in the  second panel at $t=4.88$. Similar to earlier scenarios, the vortex pair converts into a symmetric Gaussian vortex with the continuously emitted shear waves. This Gaussian vortex decay ultimately leads to its disappearance due to the quicker-emerging waves removing energy from the vortex at a faster rate.
Summery states that the merging events in VE fluids with strong circulation strengths, as opposed to HD fluids, are triggered by emerging shear waves, even for the widely positioned vortices. Higher coupling strength media accelerate the merging process, but the final vortex structure also decays faster.
\subsubsection{Medium circulation: $\Gamma_1=5$, $\Gamma_2=5$; $a_1=0.5\pi$, $a_2=0.5\pi$; $and$ $d={3\pi}$}
\label{eq_strength5_eq_size_widely}
\begin{figure}[!ht]
\centering               
\includegraphics[width=\textwidth]{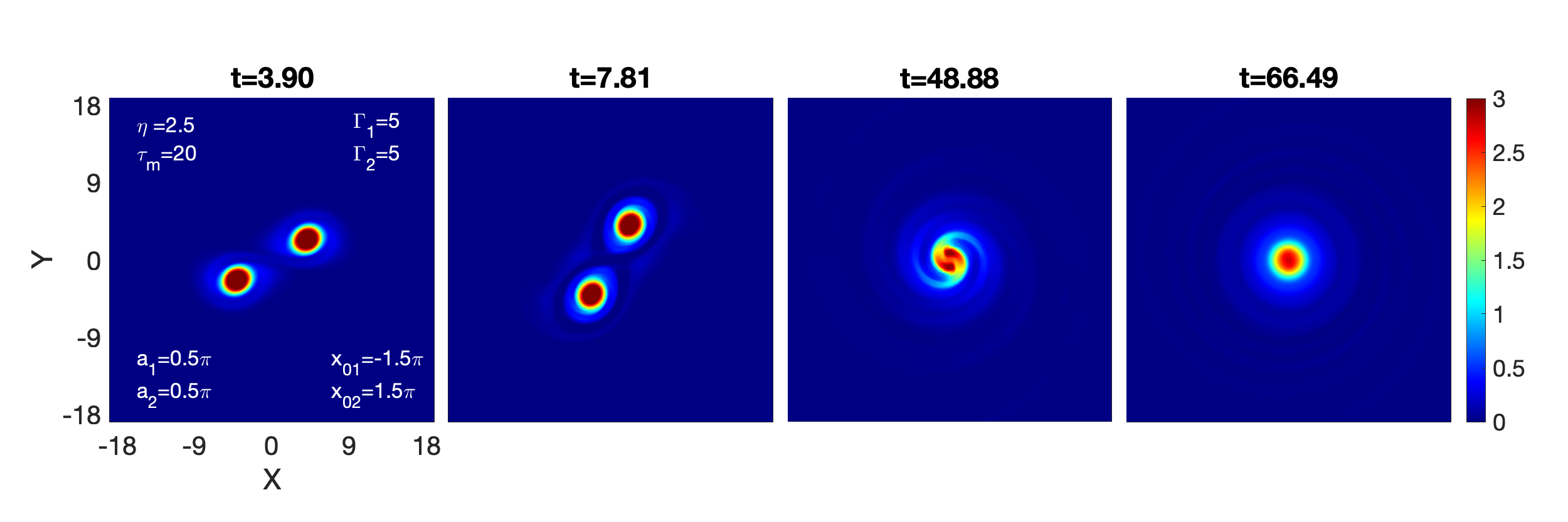}
        \caption{The time evolution of an identical vortex pair, equal circulation strength $\Gamma_0=5$, in viscoelastic fluid ($\eta=2.5$; $\tau_m=20$).}
\label{fig:figure5}
\end{figure}

\begin{figure}[!ht]
\centering               
\includegraphics[width=\textwidth]{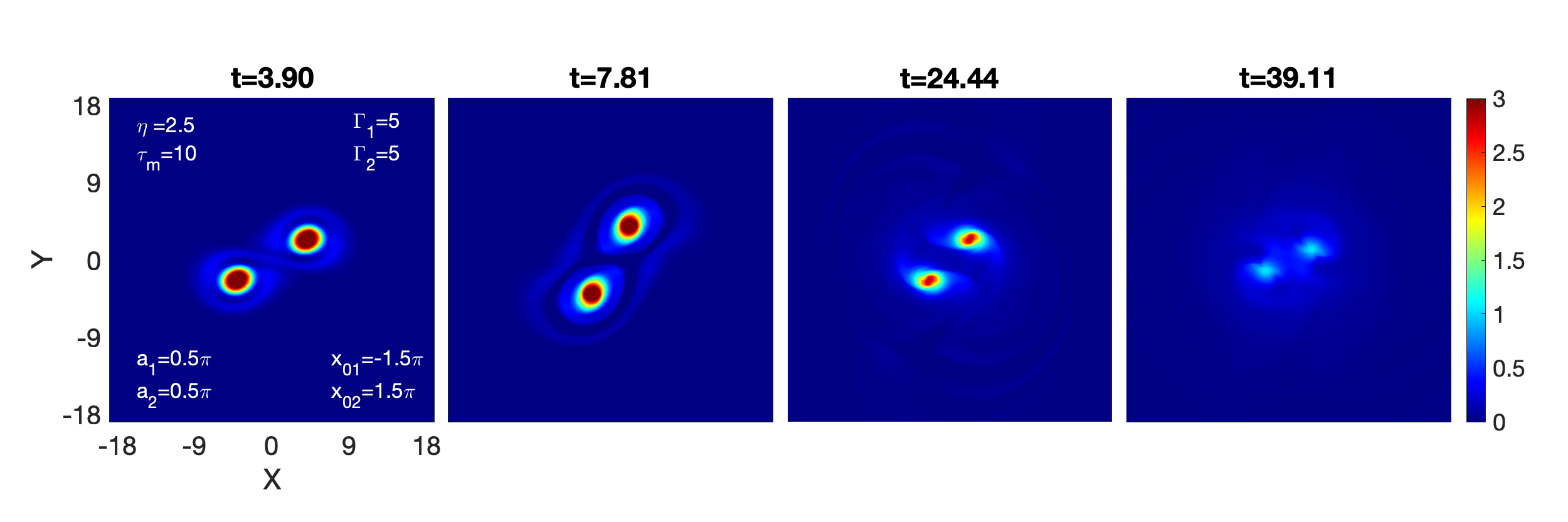}
        \caption{The time evolution of an identical vortex pair, equal circulation strength $\Gamma_0=5$, in VE fluid ($\eta= 2.5$; $\tau_m=10$).}
\label{fig:figure6}
\end{figure}

\begin{figure}[!ht]
\centering               
\includegraphics[width=\textwidth]{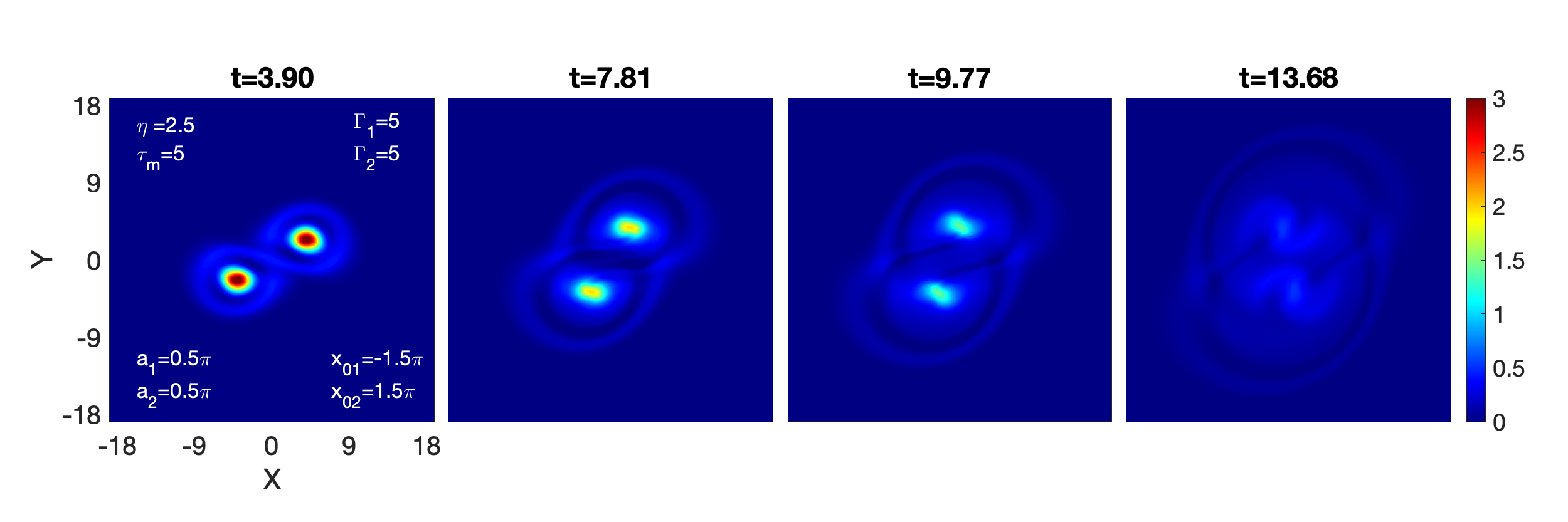}
        \caption{The time evolution of an identical vortex pair, equal circulation strength $\Gamma_0=5$, in VE fluid ($\eta=2.5$; $\tau_m=5$).}
\label{fig:figure7}
\end{figure}

\paragraph*{}
In contrast to the preceding strong circulation strength $\Gamma_0=10$, we are now considering a vortex pair with medium circulation $(\Gamma_0=5)$. For $\Gamma_0=5$, figure \ref{fig:figure5} displays the time evolution of the vorticity profile for a VE fluid with $\eta = 2.5$ and $\tau_m = 20$. When compared to strong circulation strength, a vortex pair with medium circulation strength rotates more slowly around the centriod. This implies that a slowly rotating vortex pair would have less of an impact on the shear outgoing waves, or, to put it another way, would not severely spiral the outgoing waves and the waves will assume symmetry or become circular earlier than strong circulation example. This effect is clearly evident from comparing the panel at $t = 66.73$ from figure \ref{fig:figure2} with the panel at $t = 66.49$ from figure \ref{fig:figure5} allows
one to see the faster shear waves. The depletion of energy by these outgoing waves will cause this vortex pair configuration to disappear more quickly.
In Figure \ref{fig:figure6}, the VE fluid with coupling parameters $\eta = 2.5$ and $\tau_m = 10$ is shown. This medium supports the shear waves moving at a phase velocity $v_p =\sqrt{\eta/\tau_m}$ = 0.5, which is greater than the above-mentioned scenario ($v_p =\sqrt{\eta/\tau_m}$ = 0.35; see figure \ref{fig:figure5}). The continuous shear wave emission from this vortex-pair configuration causes it to vanish before the merging event occurs.
Figure \ref{fig:figure7} displays the VE fluid with coupling parameters $\eta = 2.5$ and $\tau_m = 5$. This medium supports the shear waves moving at a phase velocity $v_p =\sqrt{\eta/\tau_m}$ = 0.71, which is greater than from both the earlier scenarios. The vortex-pair structure disappeared earlier than in the two above situations, before the merger event, as expected.
According to summery, the merging occurs for the medium with mild-strong coupling strength ($\eta$=2.5, $\tau_m$=20), where shear waves emerge slowly. However, the structures disappear even before the merging occurs as the medium gets stronger coupling strength (medium-strong: $\eta$=2.5, $\tau_m$=10; and strong or strongest: $\eta$=2.5, $\tau_m$=5). This disappearance is proportional to the coupling strength of the medium.
\subsubsection{Weak circulation: $\Gamma_1=3$, $\Gamma_2=3$; $a_1=0.5\pi$, $a_2=0.5\pi$; $and$, $d={3\pi}$}
\label{eq_strength3_eq_size_widely}
\begin{figure}[!ht]
\centering               
\includegraphics[width=\textwidth]{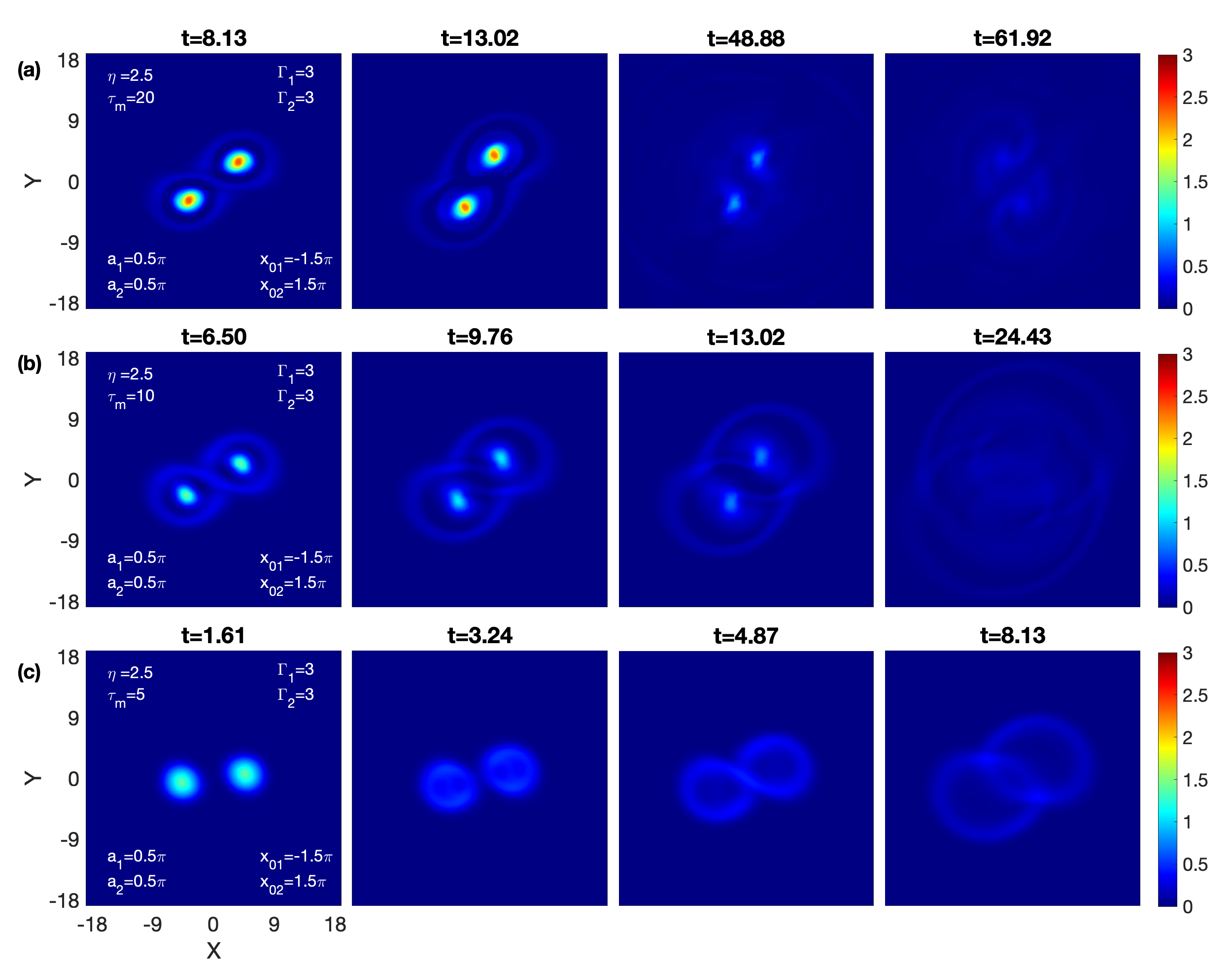}
\caption{Time evolution of an identical vortex pair with equal circulation strength $\Gamma_0=3$ in VE fluids for fixed viscous term $\eta = 2.5$; and varying coupling strength arising from changing relaxation parameter $\tau_m$= 20, 10 and 5 in (a), (b) and (c), respectively.}
\label{fig:figure8}
\end{figure}

\paragraph*{}
For weak circulation $\Gamma=3$, the figures \ref{fig:figure8}(a), \ref{fig:figure8}(b), and \ref{fig:figure8}(c) represent the VE fluids with variable relaxation parameters $\tau_m$ = 20, 10, and 5, respectively; for the fixed viscosity $\eta = 2.5$. In all three situations, the emerging waves engulfed the vortex-pair structure before the merging event occurred. This vanishing is proportional to the medium's coupling strength, as demonstrated by the comparative analysis.
 \subsubsection*{b. Closely spaced {$\left(a_0/b_0~{<}~({a_0}/{b_0})_{crit}\right)$}}
 \label{eq_strength_eq_size_closely}
\begin{figure}[!ht]
\centering               
\includegraphics[width=\textwidth]
{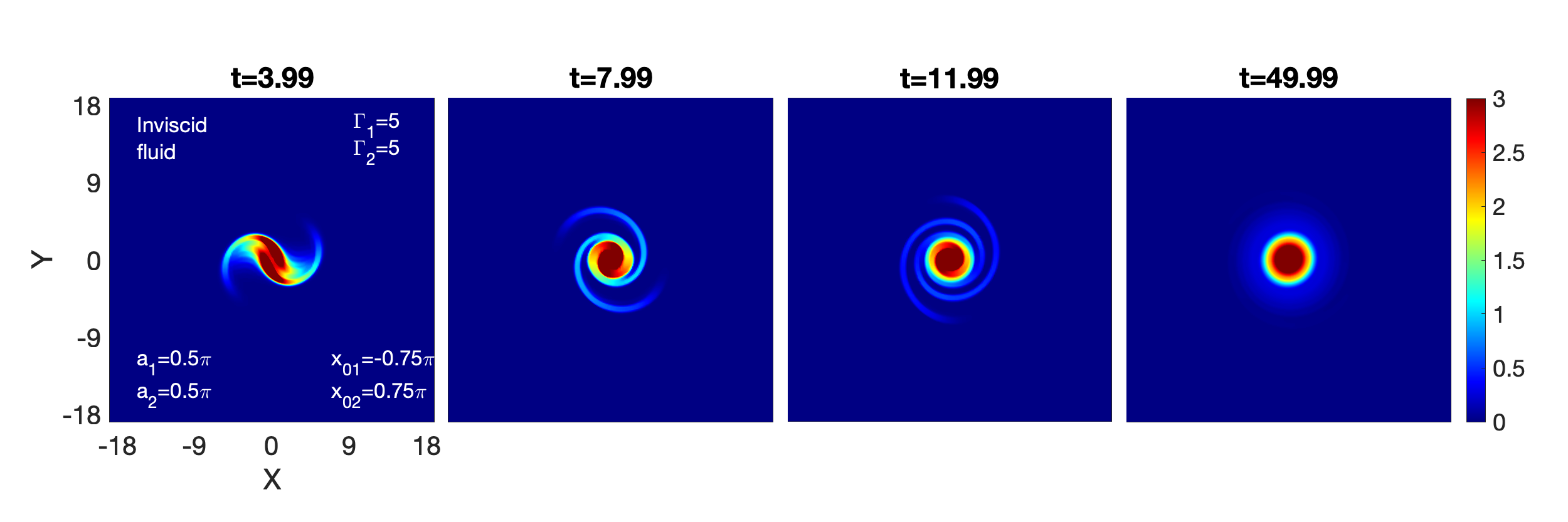}
\caption{Time evolution of  vorticity  for an inviscid HD fluid for $\Gamma_0=5$. The vortices rotate about each other due to the mutually induced velocity, no merging phenomena.}
\label{fig:figure9}
\end{figure}

\begin{figure}[!ht]
\centering               
\includegraphics[width=\textwidth]
{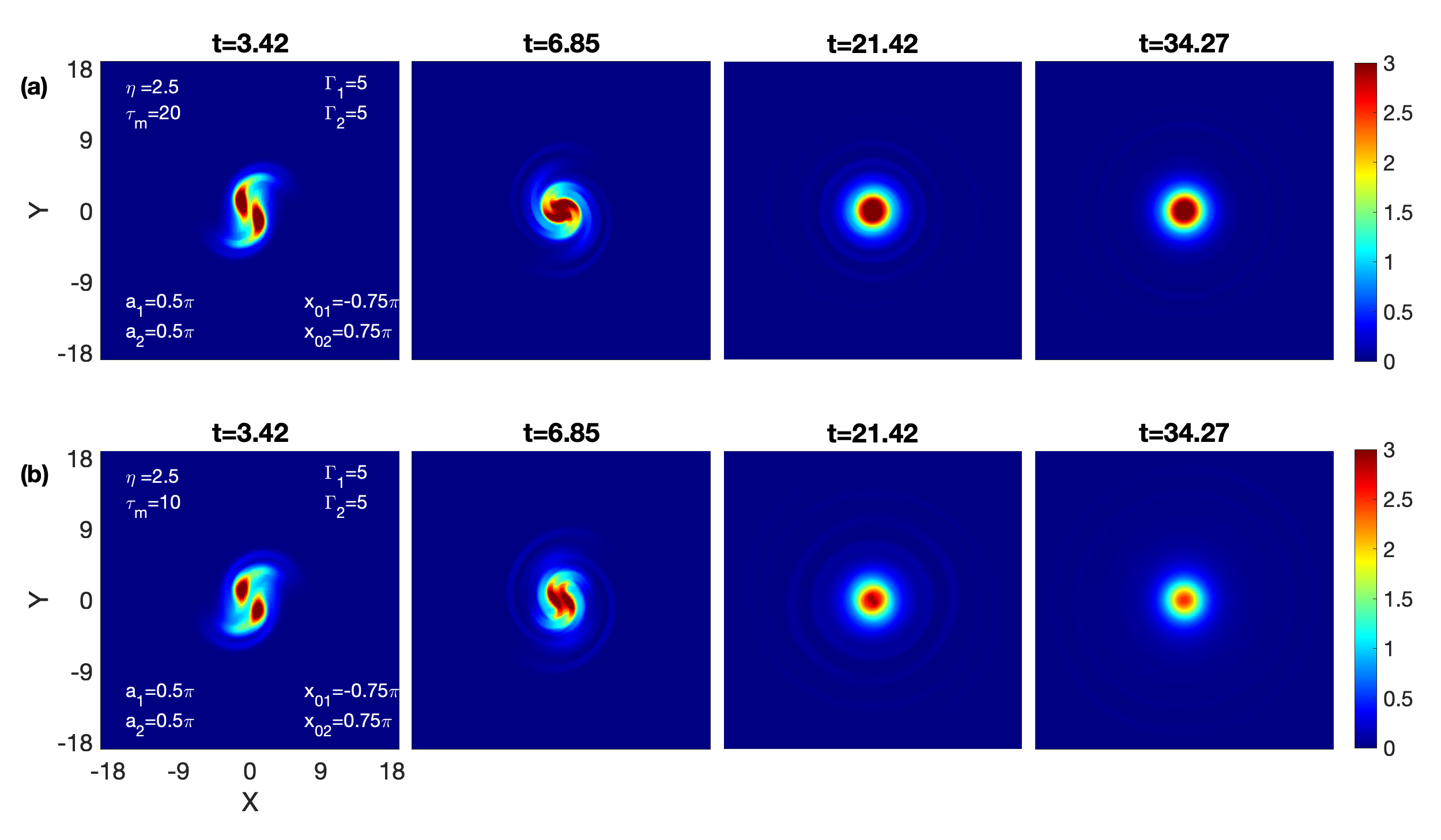}
\caption{Time evolution of an identical vortex pair with equal circulation strength $\Gamma_0=5$ in VE fluids for fixed viscous term $\eta = 2.5$; and varying coupling strength arising from changing relaxation parameter $\tau_m$= 20, and 10 in (a), and (b), respectively.}
\label{fig:figure10}
\end{figure}

\begin{figure}[!ht]
\centering               
\includegraphics[width=\textwidth]
        {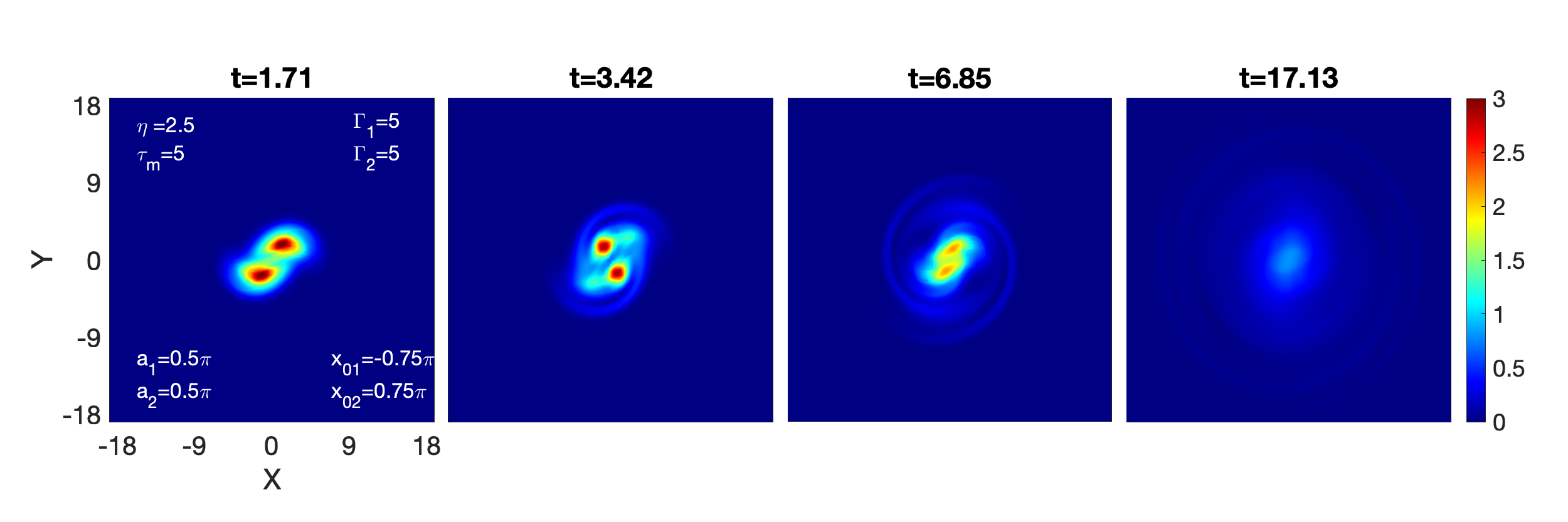}
        \caption{The time evolution of an identical vortex pair, equal circulation strength $\Gamma_0=5$, in viscoelastic fluid ($\eta=2.5$; $\tau_m=5$).}
\label{fig:figure11}
\end{figure}

Here, the vortices are centered at $({x_{01}},{x_{02}})=(-1.5a_0,1.5a_0)$, so $b_0=3a_0$. This means that the aspect ratio is $a_0/b_0=0.33$, which is initially more than the $0.29$ merger threshold, indicating that the merging phenomena is anticipated to occur.
\paragraph*{}
We have first performed the simulations for the incompressible inviscid fluids for all three circulation strengths (10, 5, and 3). The merging events is observed for all three strengths.  For $\Gamma_0=5$, the merging process in the inviscid HD fluid is shown in Figure \ref{fig:figure9}. 
Here, in SCDPs, we are merely providing a circulation strength of $\Gamma_0=5$ since the process of merging phenomenon is nearly identical to that of the previously discussed cases in widely spaced Sec. \ref{eq_strength_eq_size_widely}. The figures \ref{fig:figure10}(a) and \ref{fig:figure10}(b) represent the variable relaxation parameters $\tau_m$ = 20 and 10, respectively; for the fixed viscosity $\eta = 2.5$. The merging phenomenon is similar for both Figs. \ref{fig:figure10}(a) and \ref{fig:figure10}(b), but the envelope of the wavefront is larger in Fig. \ref{fig:figure10}(b) because of faster shear waves.
Figure \ref{fig:figure11}, we demonstrate the different stages of VE simulation, with values of $\eta= 2.5$; $\tau_m=5$ i.e. $\eta/{\tau_m}= 0.5$. The vortex-pair configuration disappears prior to the merger event at a faster rate than in the above scenarios. 
As per Summery, the merger takes place for the medium with mild-strong  coupling strength ($\eta$=2.5, $\tau_m$=20) and medium with medium-strong coupling strength ($\eta$=2.5, $\tau_m$=10). On the other hand, in a medium with strong coupling strength, the structures disappear even before the merging occurs. 
\subsection{Disparate strength and equal size: $\Gamma_1{\neq}\Gamma_2$ and ${a_1=a_2}$}
\label{uneq_strength_eq_size_widely}
\paragraph*{}
We consider a pair of disparate circulation-strength vortices ($\Gamma_1{\neq}\Gamma_2$) of equal initial radii, ${a_1=a_2=a_0=\pi/2}$ and separated by distance $b_0=6a_0$ with $({x_{01}},{x_{02}})=(-3a_0,3a_0)$. Here, the simulations are performed for three disparate strength cases (1), (2), and (3), respectively: $\Gamma_1=10$, $\Gamma_2=5$; $\Gamma_1=10$, $\Gamma_2=3$; and $\Gamma_1=5$, $\Gamma_2=3$. Once more, the suggested coupling strengths in each case are mild-strong ($\eta$=2.5, $\tau_m$=20), medium-strong ($\eta$=2.5, $\tau_m$=10), and strong or strongest ($\eta$=2.5, $\tau_m$=5).
\subsubsection{$\Gamma_1=10$, $\Gamma_2=5$; $a_1=0.5\pi$, $a_2=0.5\pi$; $and$ $d={3\pi}$}
\label{uneq_strength_g10g5_eq_size_widely}
\begin{figure}[!ht]
\centering               
\includegraphics[width=\textwidth]
        {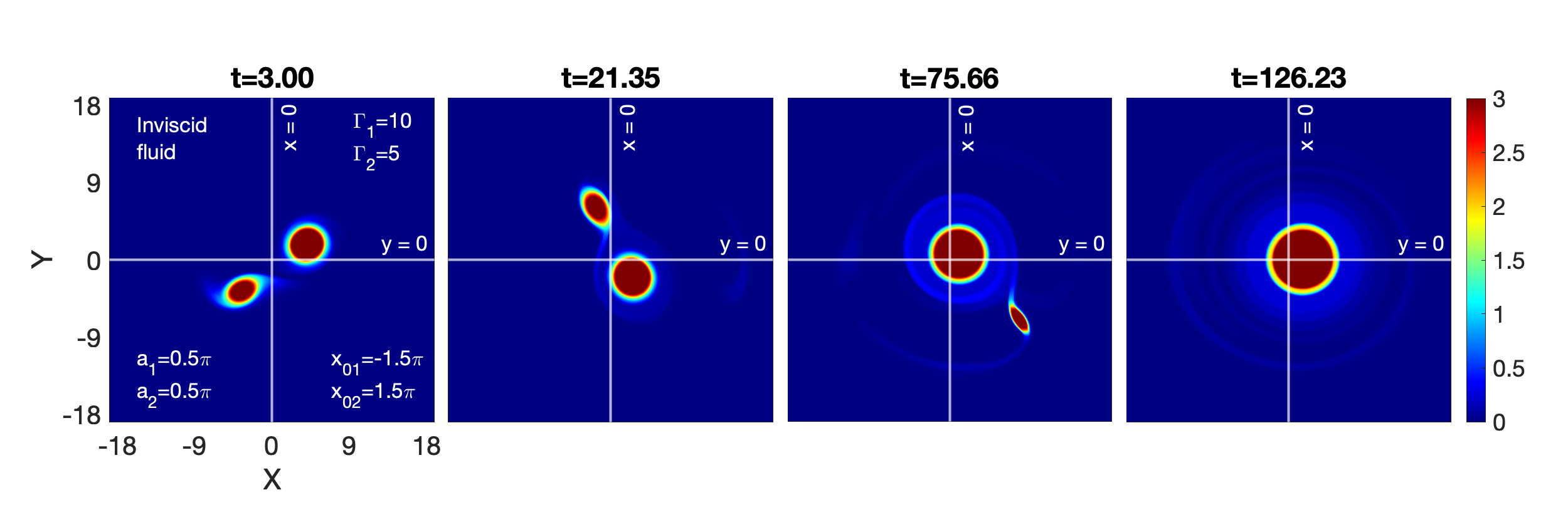}        \caption{Merging between two like sign disparate strength vortices in inviscid HD fluid with $\Gamma_1=10$ and $\Gamma_2=5$ .}
\label{fig:figure12}
\end{figure}

\begin{figure}[!ht]
\centering               
\includegraphics[width=\textwidth]
        {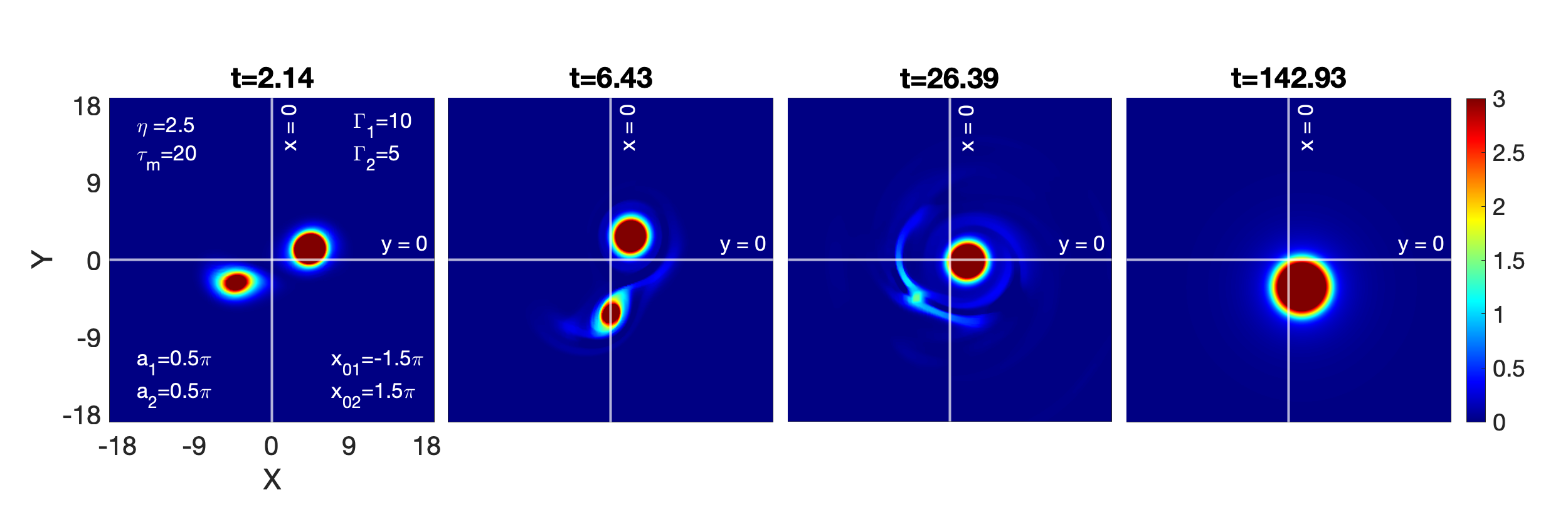}
        \caption{Merging between two like sign disparate strength vortices in viscoelastic fluids for fixed viscous term $\eta = 2.5$; and  $\tau_m$= 20 with $\Gamma_1=10$ and $\Gamma_2=5$. Notice the periodic deformation from a circle to a nearly elliptical shape of the weaker vortex as well as the coherent rotation of the major axis. The larger vortex experienced almost no deformation.}
 \label{fig:figure13}
\end{figure}

\begin{figure}[!ht]
\centering               
\includegraphics[width=\textwidth]
        {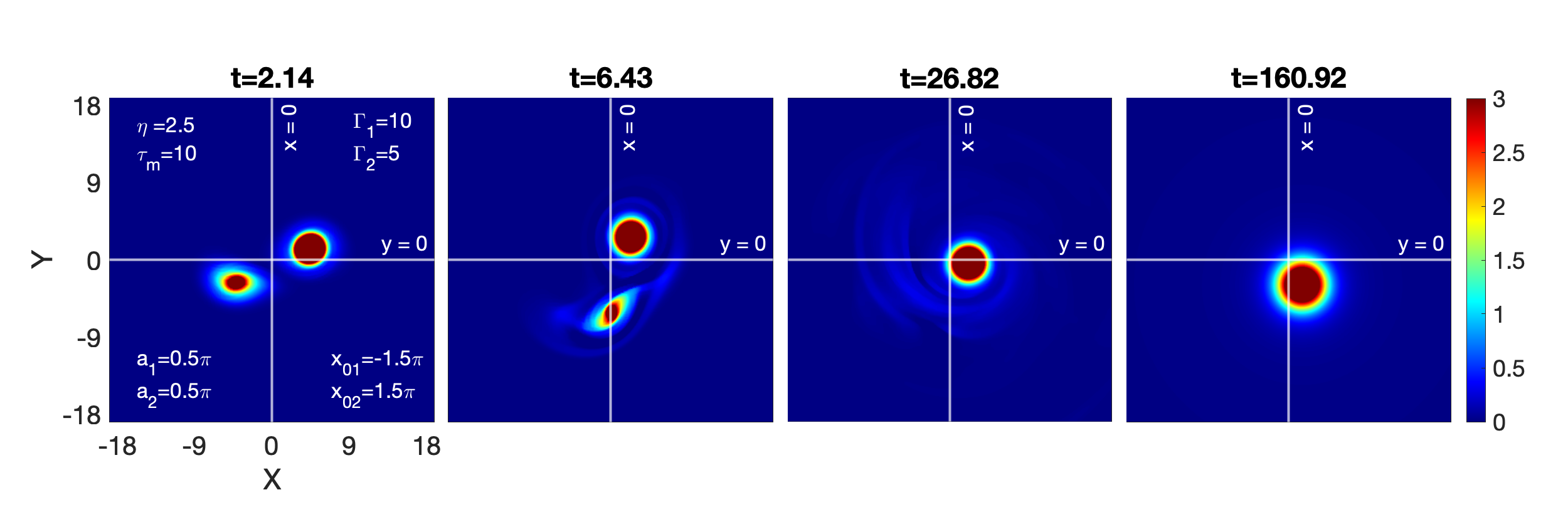}
        \caption{Merging between two like sign disparate strength vortices in viscoelastic fluids for fixed viscous term $\eta = 2.5$; and  $\tau_m$= 10 with $\Gamma_1=10$ and $\Gamma_2=5$.}
        \label{fig:figure14}
\end{figure}

\begin{figure}[!ht]
\centering               
\includegraphics[width=\textwidth]
{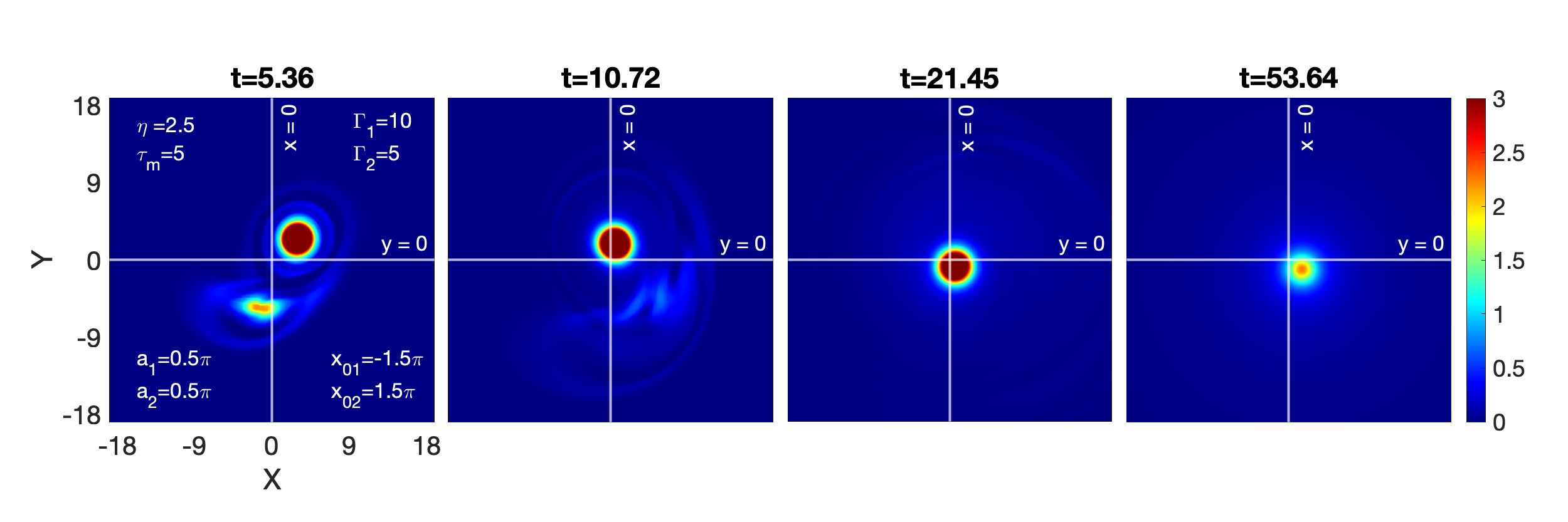}
\caption{Merging between two like sign disparate strength vortices in viscoelastic fluids for fixed viscous term $\eta = 2.5$; and  $\tau_m$= 20 with $\Gamma_1=10$ and $\Gamma_2=5$. }
    \label{fig:figure15}
\end{figure}

\paragraph*{}
In this case, we consider two different circulation-strength vortices, with $\Gamma_1=10$ representing a stronger vortex and $\Gamma_2=5$ representing a weaker vortex. The strength difference, $d{\Gamma}=\Gamma_1-\Gamma_2=5$. The figure \ref{fig:figure12} illustrates the temporal evolution of such vorticies in inviscid fluids. When vortices are not equal, they do not distort at the same rate. The deformation rate of weaker vortex is higher relative to the stronger vortex. As two vortices revolve around one another, the weaker vortex undergoes periodic deformations. During this process, the weaker vortex changes from its original circular shape to nearly elliptical, while the stronger vortex remains almost circular. After undergoing several rotational deformations, the elliptical vortex gradually disappeared and left a spiral structure surrounding the stronger vortex.  The center of the final vortex with spiral structure has moved slightly off center. Such observations are clearly evident in figure \ref{fig:figure12}.
The figure \ref{fig:figure13} displays how these vorticies ($\Gamma_1=10$, $\Gamma_2=5$) interact over time in a VE fluid with $\eta = 2.5$ and $\tau_m$= 20. Similar to inviscid HD fluid (see figure \ref{fig:figure12}), the stronger vortex appears to be consistent throughout the interaction and deviates somewhat from the center. During this process, in the beginning, the weaker vortex changes from its original circular shape to nearly elliptical, while later it gets distorted form. Distorted form is because the weaker vortex is constantly releasing shear waves and also interacting with the stronger shear wave coming from the rotating stronger vortex. It seems that in the end, shear waves engulf the weaker vortex before it combines with the stronger vortex. In contrast to the HD fluid (see figure \ref{fig:figure12}), the weaker vortex does not vanish smoothly and there is no spiral structure surrounding the final vortex.
Figure \ref{fig:figure14} shows VE fluid with parameters $\eta = 2.5$ and $\tau_m = 10$. This medium supports the shear waves traveling at a phase velocity of 0.5, which is higher than the 0.35 situation described above (see figure \ref{fig:figure13}). As a result, here, the weaker vortex disappears sooner than figure \ref{fig:figure13}.
A new VE fluid with a stronger coupling strength $(\eta = 2.5, \tau_m = 5.0)$ than the earlier scenarios is shown in figure \ref{fig:figure15}. This medium supports the TS waves at a phase velocity $v_p =\sqrt{\eta/\tau_m}$ = 0.71, which is faster than above discussed scenarios. Here, as a result, the weaker vortex vanishes earlier than in the previously described circumstances. Additionally, because the shear waves are emerging faster and extracting energy from the vortex at a higher rate, the final smooth vortex eventually disappears.
In summary, for both the HD and VE fluids, we observe that the final vortex gets slightly out of center and the weaker vortex disappears. Because shear waves are present in VE fluids, this happens more quickly than in HD fluid. In VE fluids, disappearance of the weaker vortex is proportional to the coupling strength of the medium. Moreover, we also observe the stronger vortex vanishing for the strong coupling strength medium in less than half the time in other two scenarios.

\subsubsection{$\Gamma_1=10$, $\Gamma_2=3$; $a_1=0.5\pi$, $a_2=0.5\pi$; $and$ $d={3\pi}$}
\label{uneq_strength_g10g3_eq_size_widely}
\begin{figure}[!ht]
\centering               
\includegraphics[width=\textwidth]
        {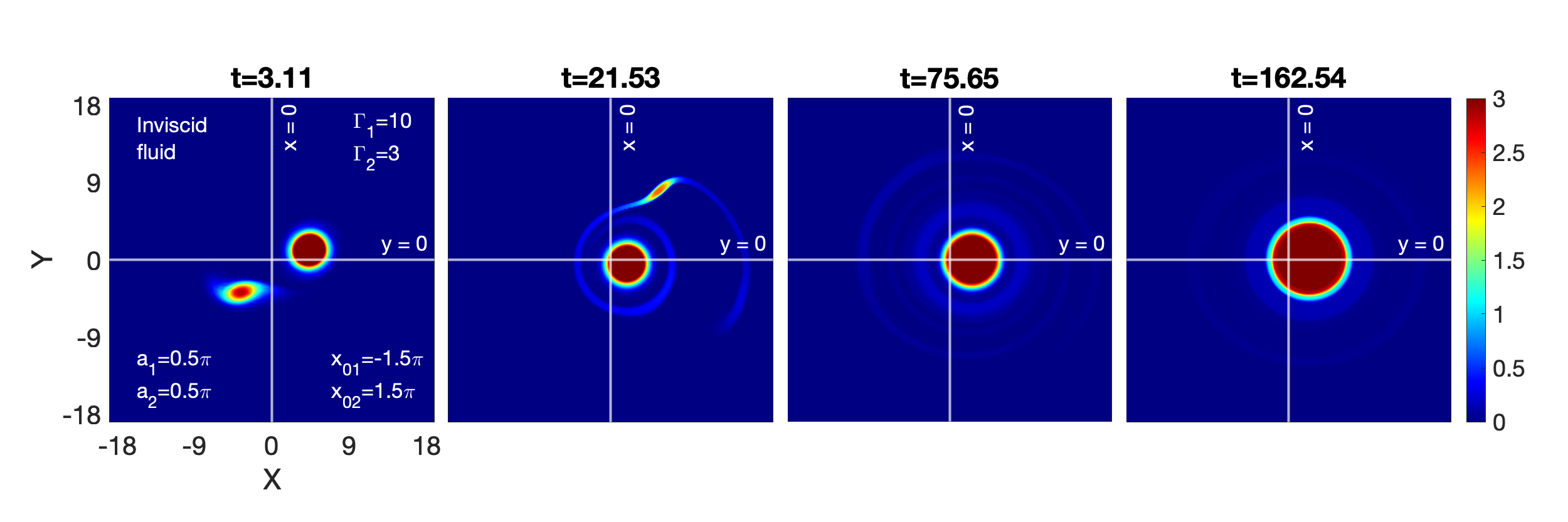}
\caption{Merging between two like sign disparate strength vortices in inviscid HD fluid with $\Gamma_1=10$ and $\Gamma_2=3$ .}
\label{fig:figure16}
\end{figure}

\begin{figure}[!ht]
\centering               
\includegraphics[width=\textwidth]
{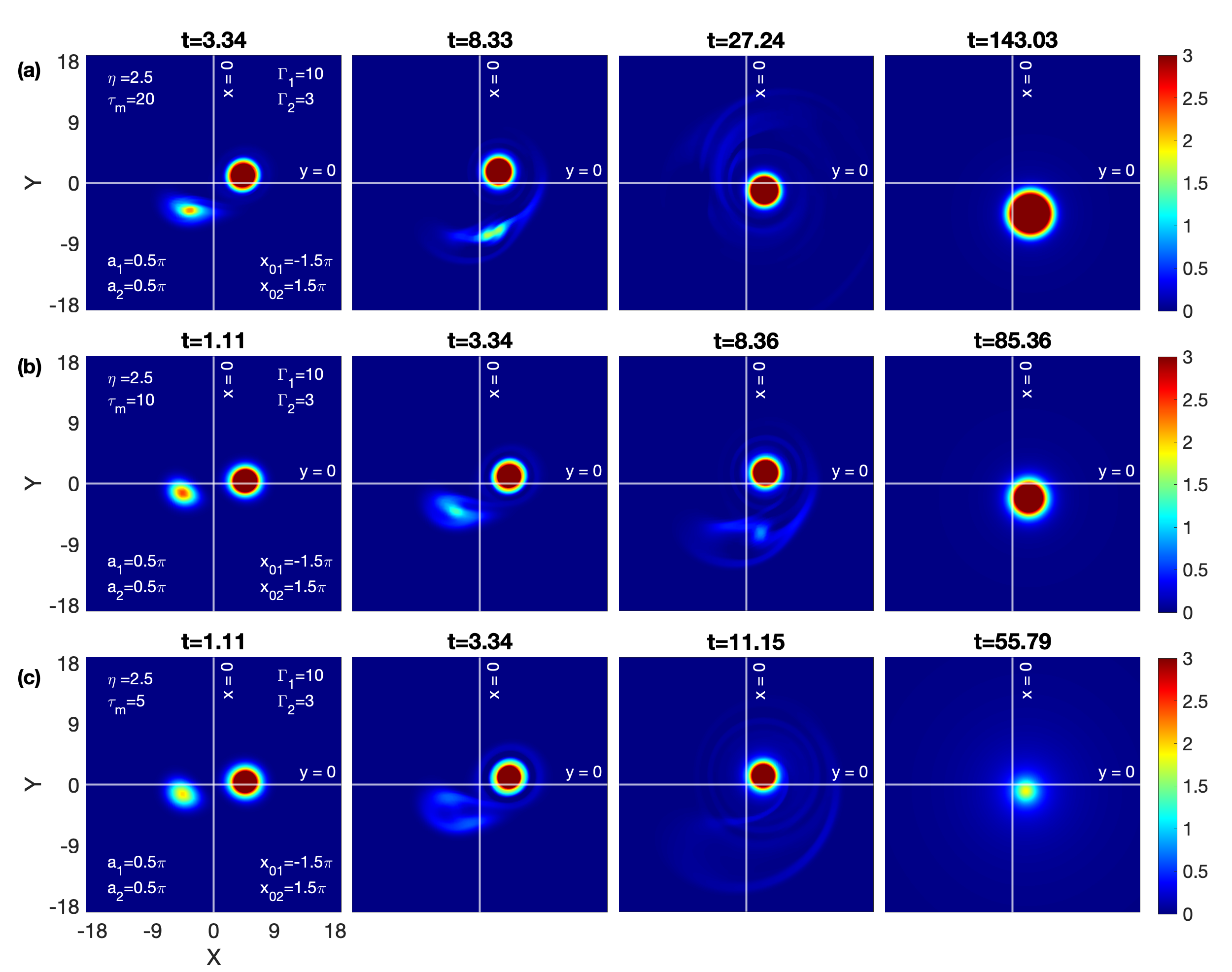}
\caption{Time evolution of an identical vortex pair with equal circulation strength $\Gamma=3$ in VE fluids for fixed viscous term $\eta = 2.5$; and varying coupling strength arising from changing relaxation parameter $\tau_m$= 20, 10 and 5 in (a), (b) and (c), respectively.}
\label{fig:figure17}
\end{figure}

\paragraph*{}
In this case, we lower the strength of the weaker vortex from 5 to 3 compared to the previous case. $d{\Gamma}=\Gamma_1-\Gamma_2=7$. So, we have two vortices with $\Gamma_1=10$ (stronger vortex) and $\Gamma_2=3$ (weaker vortex). Figure \ref{fig:figure16} illustrates the interaction process between these vorticies, which is similar to the previous fluid instance (see figure \ref{fig:figure12}) but with the weaker vortex that vanishes sooner.
For VE fluids with fixed viscosity $\eta = 2.5$, the varied relaxation parameters $\tau_m$ = 20, 10, and 5 are represented in figures \ref{fig:figure17}(a), \ref{fig:figure17}(b), and \ref{fig:figure17}(c), respectively. The comparative analysis reveals that the disappearing of a weaker vortex is proportional to the coupling strength of the medium in all three cases.
To sum up, in both the HD and VE fluids, the weaker vortex ($\Gamma=3$) vanishes more quickly than in the earlier examples ($\Gamma=5$) covered in Sec. \ref{uneq_strength_g10g5_eq_size_widely}. This is also because, in addition to the  lower the strength of the weaker vortex, the strength differential $d{\Gamma}=7$ is larger in this case than the earlier $d{\Gamma}=5$.

\subsubsection{$\Gamma_1=5$, $\Gamma_2=3$; $a_1=0.5\pi$, $a_2=0.5\pi$; $and$~$d={3\pi}$}
\label{uneq_strength_g5g3_eq_size_widely}
\begin{figure}[!ht]
\centering               
\includegraphics[width=\textwidth]
{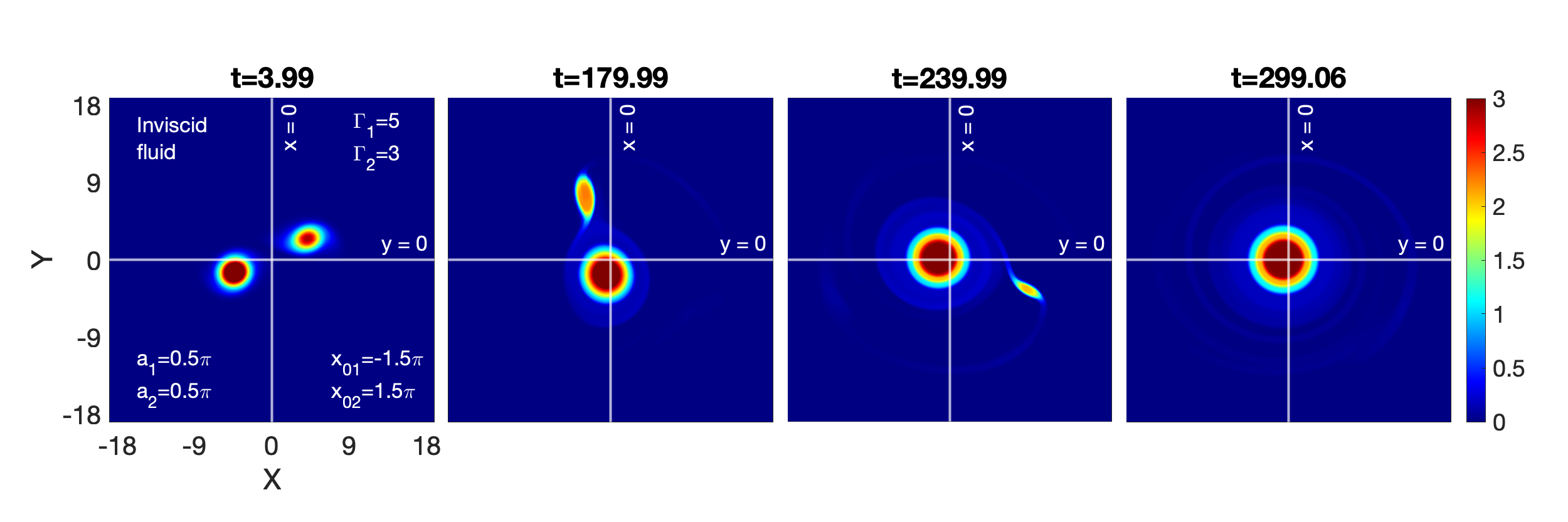}
\caption{Merging between two like sign disparate-size vortices in inviscid HD fluid with $a_1=10$ and $a_2=5$ .}
\label{fig:figure18}
\end{figure}

\begin{figure}[!ht]
\centering               
\includegraphics[width=\textwidth]
        {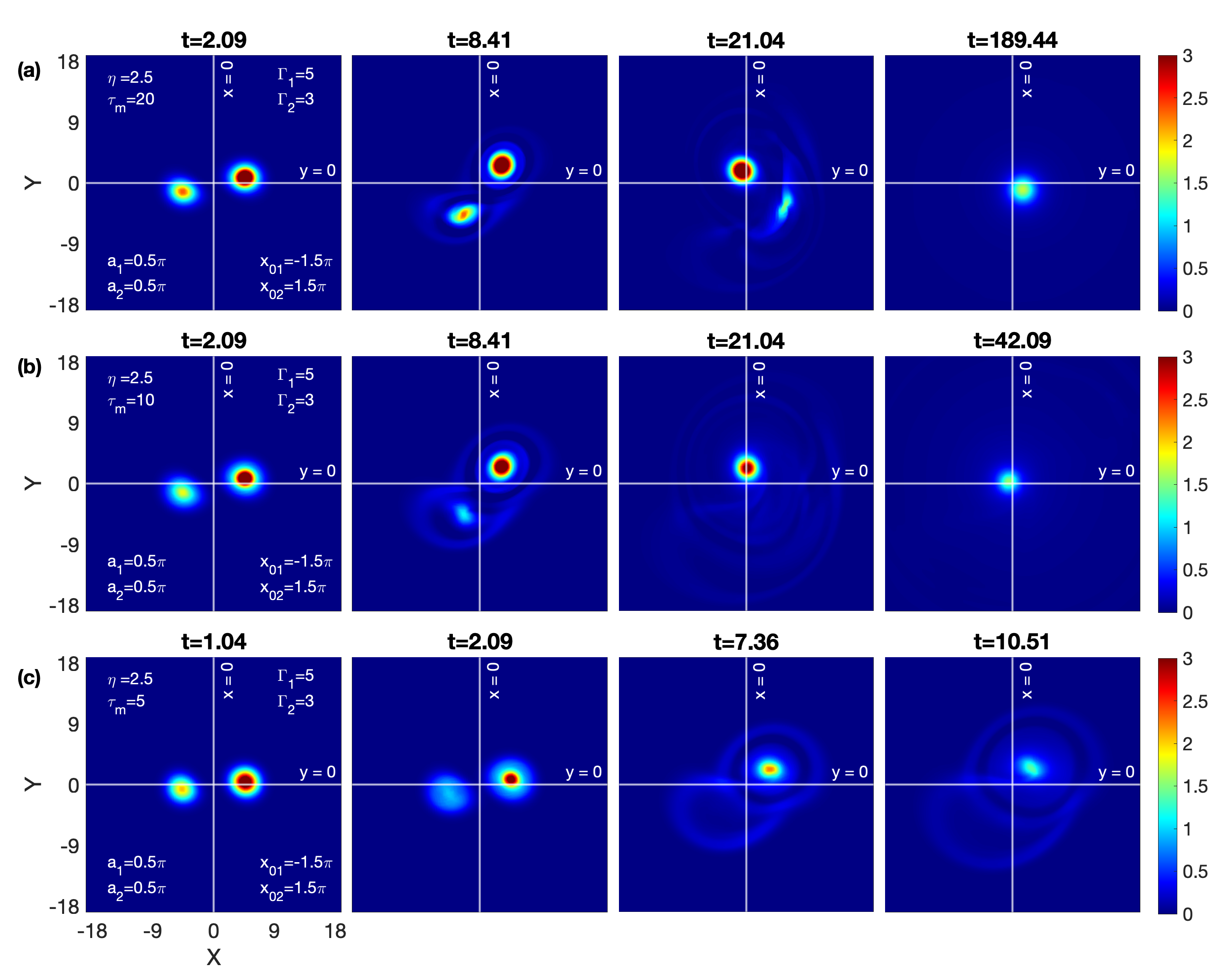}   
\caption{Time evolution of an identical vortex pair with equal circulation strength $\Gamma=3$ in VE fluids for fixed viscous term $\eta = 2.5$; and varying coupling strength arising from changing relaxation parameter $\tau_m$= 20, 10 and 5 in (a), (b) and (c), respectively.}
\label{fig:figure19}
\end{figure}

\paragraph*{}
Compared to the prior case (Sec. \ref{uneq_strength_g10g3_eq_size_widely}), we reduce the strength of the stronger vortex from 10 to 5 in this case. So, here, we consider two distinct circulation-strength vortices: $\Gamma_1=5$ (stronger vortex) and $\Gamma_2=3$ (weaker vortex). $d{\Gamma}=\Gamma_1-\Gamma_2=2$. Figure \ref{fig:figure18} shows the temporal evolution of these vorticies in an inviscid fluid. The interaction process between these vorticies is depicted in figure \ref{fig:figure18} and is the same as that of the preceding fluid instance (see figure \ref{fig:figure12}). However, in this case, the weaker vortex survives longer because there is less of a strength differential ($d{\Gamma}=2$) between the vortices.
Figures \ref{fig:figure19}(a), \ref{fig:figure19}(b), and \ref{fig:figure19}(c)] illustrate the various relaxation parameters $\tau_m$ = 20, 10, and 5 for VE fluids with a fixed viscosity $\eta = 2.5$. A slightly deformed filament shape of the weaker vortex can be seen in the mild-strength fluid as shown in figure \ref{fig:figure19}(a). In the other two examples (figures \ref{fig:figure19}(b) and \ref{fig:figure19}(c)), the dominance of the shear waves can be seen as a result of which no filament shape of the weaker vortex occurring and the vortex pair disapearing at a faster rate. The comparative analysis demonstrates that in all three cases, the disappearance of both vortices is proportional to the coupling strength of the medium.
In conclusion, since both vortices in VE fluids are weaker than in the preceding cases, an envelope of shear waves encircling the vortex pair is seen in figures \ref{fig:figure19}(b) and \ref{fig:figure19}(c).
\subsection{Equal strength and disparate size: $\Gamma_1{=}\Gamma_2$ and $a_1{\neq}a_2$}
\label{eq_strength_uneq_size_widely_A}
\paragraph*{}
Here, we consider a vortex pair of disparate-size ${a_1{\neq}a_2}$ of equal strength $\Gamma_1{=}\Gamma_2=10$ and separated by distance $b_0=6a_0$ with $({x_{01}},{x_{02}})=(-3a_0,3a_0)$. Two distinct radii vortices are taken into consideration; a larger vortex is represented by $a_1=0.6$, while a smaller vortex is $a_2=0.4$.
\subsubsection{$\Gamma_1=10$, $\Gamma_2=10$; $a_1=0.6\pi$, $a_2=0.4\pi$; $and$ $d={3\pi}$}
\label{eq_strength10_uneq_size_widely}
\begin{figure}[!ht]
\centering               
\includegraphics[width=\textwidth]{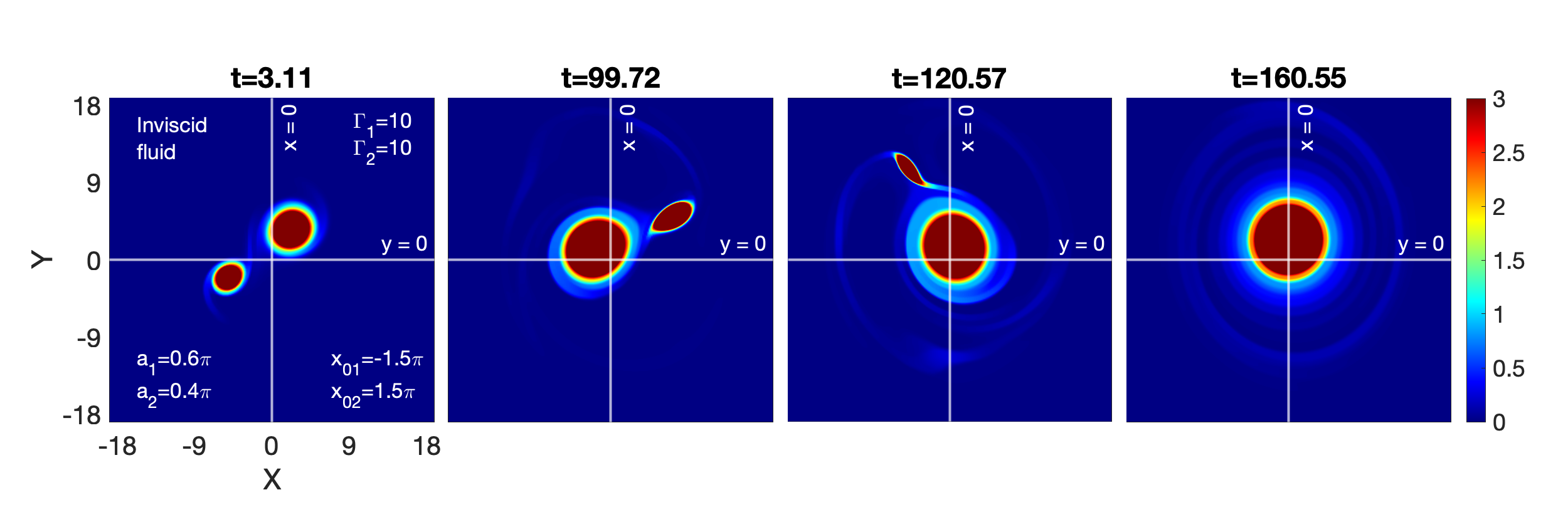}
\caption{Merging between two like sign disparate-size vortices in inviscid HD fluid with $a_1=0.6$ and $a_2=0.4$ .}
\label{fig:figure20}
\end{figure}

\begin{figure}[!ht]
\centering               
\includegraphics[width=\textwidth]
        {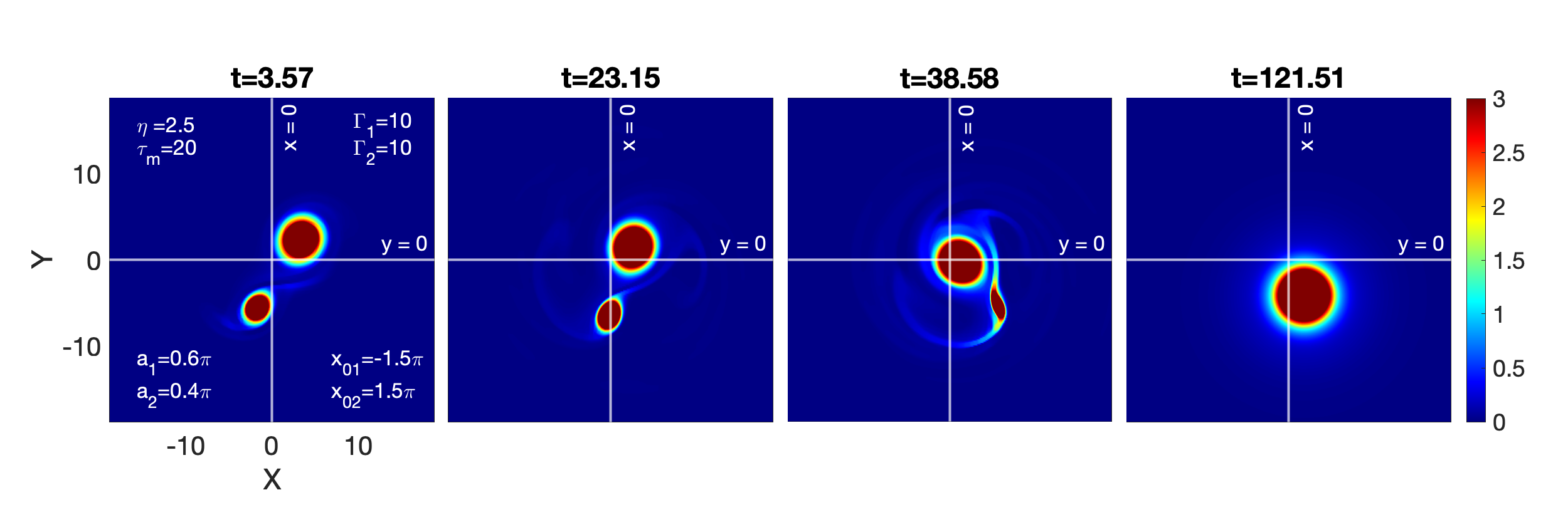}
        \caption{Merging between two like sign disparate-size vortices in VE fluid with $\eta = 2.5$, $\tau_m$ = 20 and $a_1=0.6$ and $a_2=0.4$ .}
        \label{fig:figure21}
\end{figure}

\begin{figure}[!ht]
\centering 
\includegraphics[width=\textwidth]
        {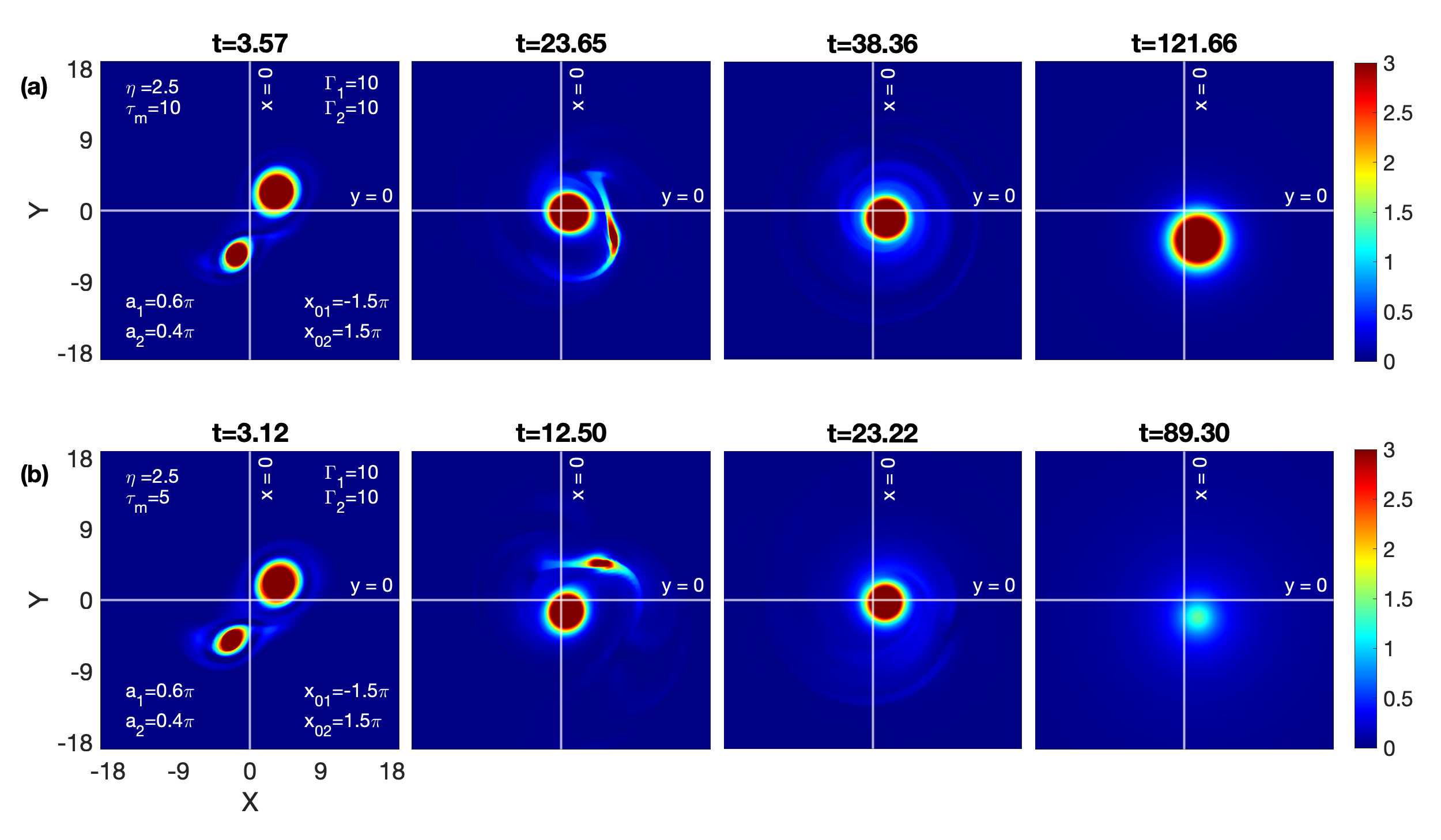}
        \caption{Merging between two like sign disparate stize vortices in viscoelastic fluids for fixed viscous term $\eta = 2.5$; and  $\tau_m$= 10 with $a_1=0.6{\pi}$ and $a_2=0.4{\pi}$.}
        \label{fig:figure22}
\end{figure}

\paragraph*{}
Figure \ref{fig:figure20} displays the time evolution of this vortex pair in an inviscid fluid. These different-size vortices have different rates of distortion. In comparison to the larger vortex, the smaller vortex deforms at a faster rate. When two vortices revolve around each other, the larger vortex stays nearly circular while the smaller vortex becomes almost elliptical. Following multiple rotational deformations, the elliptical vortex gradually disappeared and left a spiral structure around the stronger vortex.  The center of the final vortex has shifted slightly off center. These observations are quite visible in figure \ref{fig:figure20}.
The figure \ref{fig:figure21} displays a mild-strong VE fluid with $\eta = 2.5$ and $\tau_m$= 20. As in the case of inviscid HD fluid (figure \ref{fig:figure20}), the larger vortex appears to be somewhat centrally deviating throughout the encounter.The smaller vortex undergoes a major transformation during this process, initially taking on a nearly elliptical shape but later taking on a distorted form. The reason for the distorted form is its interaction with the shear waves coming from the rotating larger vortex, and the smaller vortex by itself also constantly releases shear waves, which makes it further weaker. Ultimately, the smaller vortex appears to be engulfed by shear waves before to merging with the larger vortex. Unlike the HD fluid (figure \ref{fig:figure20}), there is no spiral structure encircling the final vortex.
The figures \ref{fig:figure22}(a) and \ref{fig:figure22}(b) represent the variable relaxation parameters $\tau_m$ = 10 and 5, respectively; for the fixed viscosity $\eta = 2.5$. The comparative analysis reveals that the disappearing of a smaller vortex is proportional to the coupling strength of the medium. In figure \ref{fig:figure15}, medium supports the TS waves at a phase velocity $v_p =\sqrt{\eta/\tau_m}$ = 0.71, which is faster than \ref{fig:figure21} and \ref{fig:figure22}(a). Here, the final smooth vortex eventually vanishes because the shear waves are emerging faster and removing energy from the vortex at a higher rate.
\subsubsection{$\Gamma_1=5$, $\Gamma_2=5$; $a_1=0.6\pi$, $a_2=0.4\pi$; $and$~$d={3\pi}$}
\label{eq_strength5_uneq_size_widely}
\begin{figure}[!ht]
\centering               
\includegraphics[width=\textwidth]{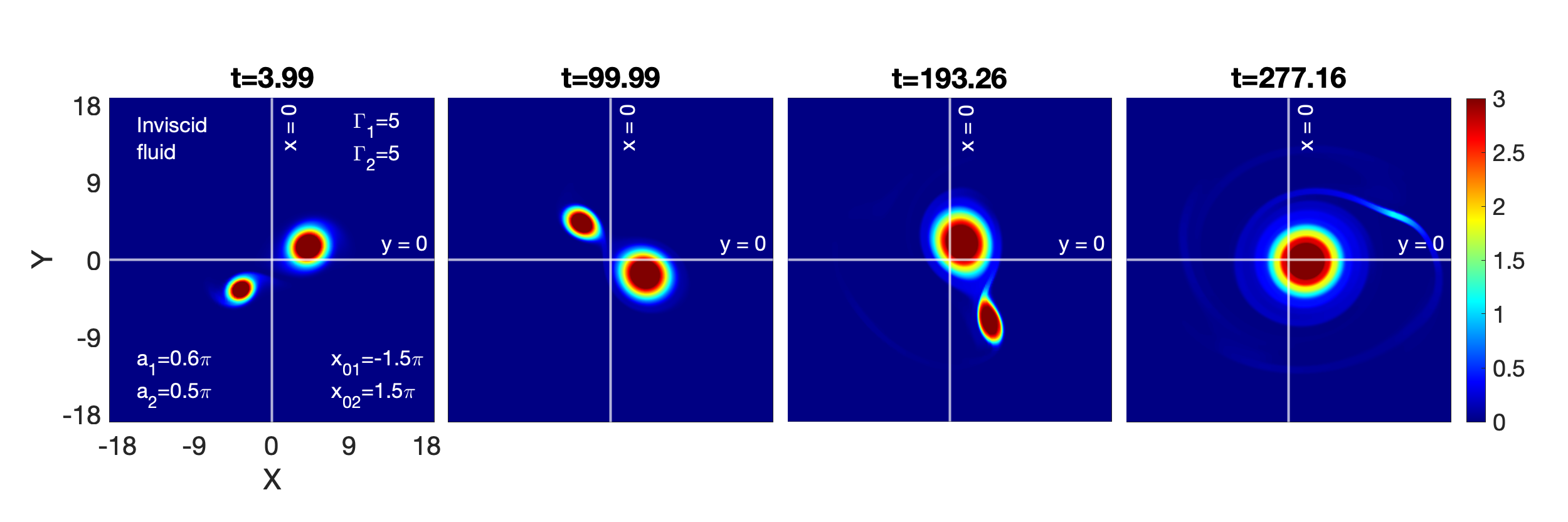}
\caption{Merging between two like sign disparate-size vortices in inviscid HD fluid with $a_1=10$ and $a_2=7.5$.}
\label{fig:figure23}
\end{figure}

\begin{figure}[!ht]
\centering 
\includegraphics[width=\textwidth]
{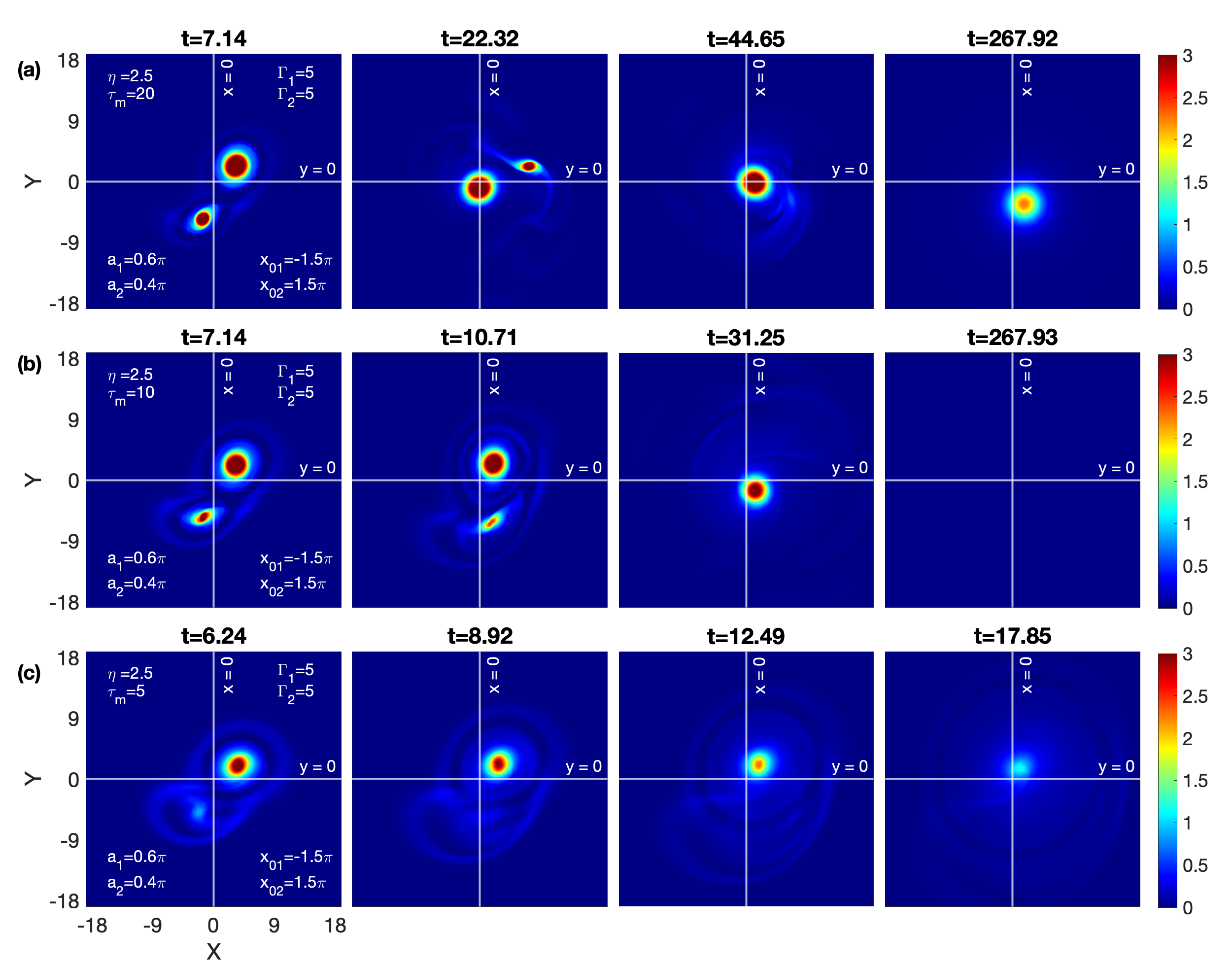}
\caption{Merging between two like sign disparate stize vortices in viscoelastic fluids for fixed viscous term $\eta = 2.5$; and  $\tau_m$= 10 with $a_1=0.6$ and $a_2=0.4$ .}
\label{fig:figure24}
\end{figure}

\paragraph*{}
Figure \ref{fig:figure23} displays the time evolution of this vortex pair of strength $\Gamma=5$ in an inviscid fluid. Here, the merging steps resemble the figure \ref{fig:figure20} scenario that was previously addressed but at a faster rate.
The comparative analysis of different scenarios in figure \ref{fig:figure24} with the respective scenarios of VE fluid discussed in the previous case for $\Gamma=10$ also shows the merging steps resemble each other but more quickly.
In summary, the merging process for different diameters bears a striking resemblance to the vortices with different strengths, where the bigger vortex acts like a stronger vortex and the smaller vortex acts like a weaker vortex.
 \subsection{Numerical validation of the conserved  equation and quantity $W$}
 \label{simulation_Conserved_quantity}
\paragraph*{}
We will now select a few of the aforementioned scenarios to examine the accuracy with which the Poynting-like conservation theorem is satisfied. We take into account a common VE fluid with parameters $\eta=2.5$ and $\tau_m=10$ in all cases. Thus, the respective circulation strengths of the vorticies and their core diameters are the varied quantities. In every vortex pair evolution plot, we draw a circle with a radius of $4{\pi}$ units that is centered at (0,0) or shares the same center as the vortex pair. Within this circular boundary, the various transport quantities are computed in order to quantify the impact of different transport processes on the evolution of W, as expressed by the integral equation (\ref{eq:integral_equ}). Additionally, we have lowered the range of colorbar in vorticity contour plots to facilitate the visualization of features like shear waves.
\subsubsection{$\Gamma_1=5$, $\Gamma_2=5$; $a_1=0.5\pi$, $a_2=0.5\pi$; $and$ $d={3\pi}$}
\label{eq_strength5_eq_size_widely_STP}
\begin{figure}[!ht]
\centering 
\includegraphics[width=\textwidth]        {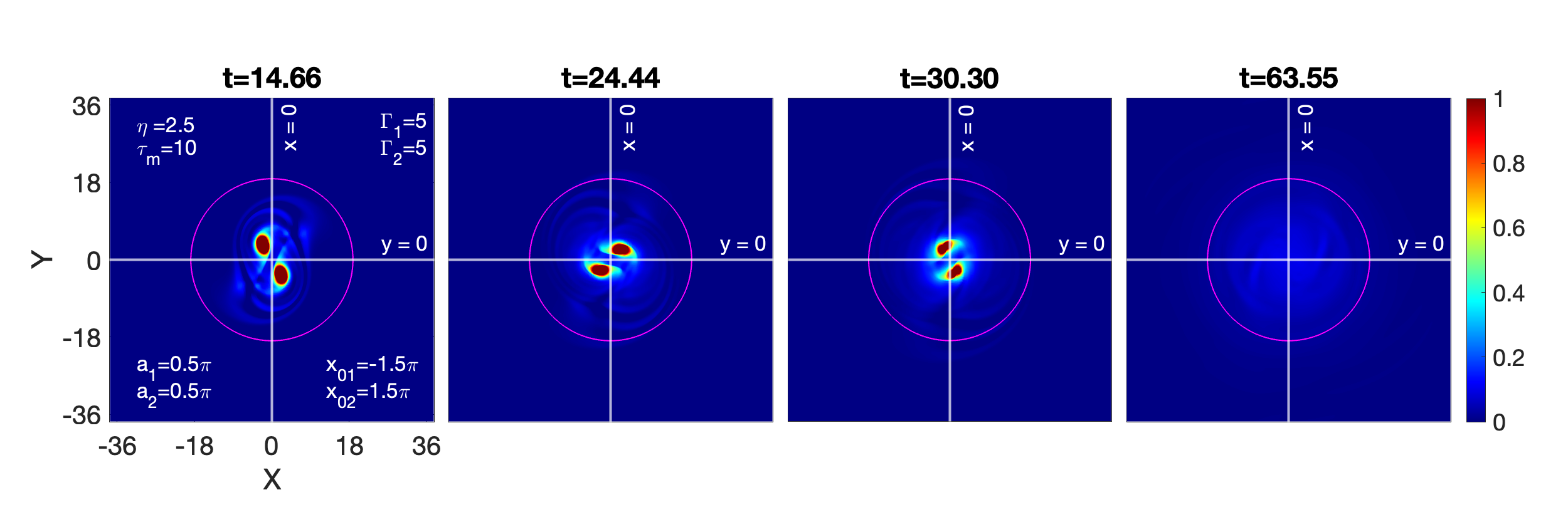}
\caption{Re-plot the figure~\ref{fig:figure6}. The time evolution of an identical vortex pair, equal circulation strength $\Gamma_0=5$, in viscoelastic fluid ($\eta=2.5$; $\tau_m=10$). The various transport quantities are calculated across a circular local volume element that is inside the circumference.}
\label{fig:figure25}
\end{figure}

\begin{figure}[!ht]
\centering 
\includegraphics[width=\textwidth]
{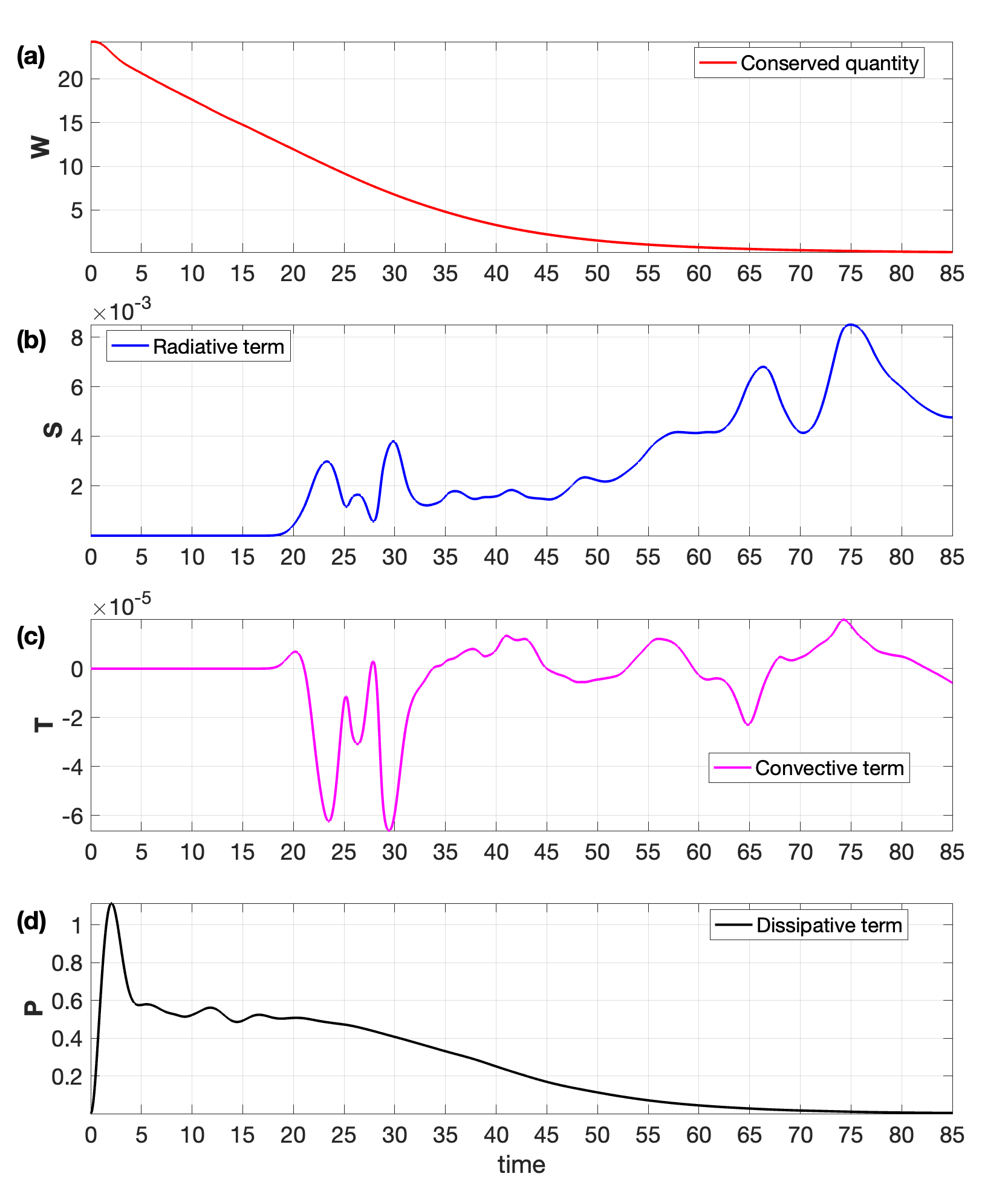}
\caption{The subplot (a) represents the change in W by wave emission. The contribution of convection is shown in subplot (b). The role of dissipating term is shown in subplot (c), which is observed to be finite.}
\label{fig:figure26}
\end{figure}

\begin{figure}[!ht]
\centering 
\includegraphics[width=\textwidth]
{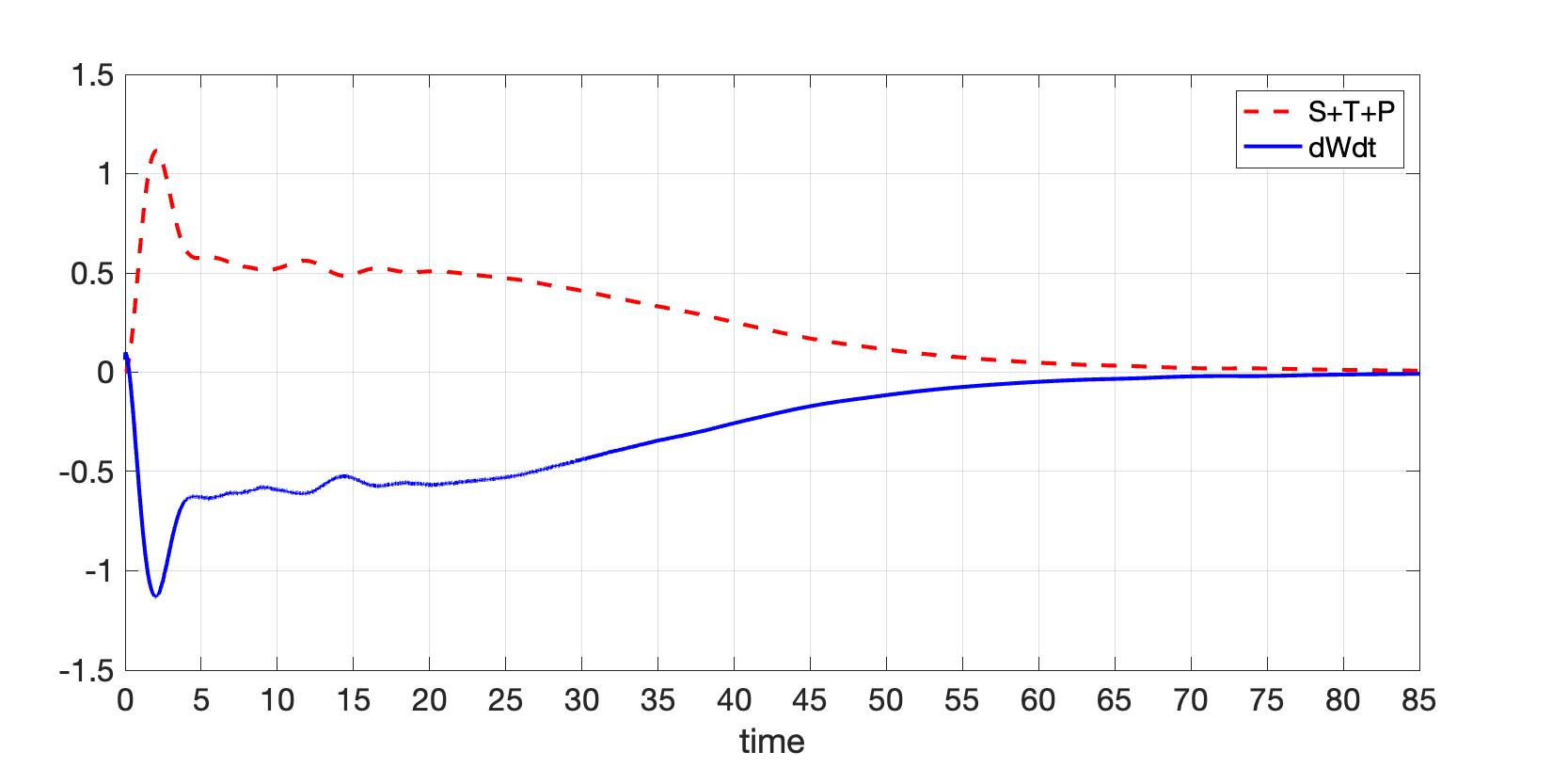}
\caption{Time derivative of conserved quantity W (–) is the mirror image to the total sum ($\bf{S+T+P}$) of all remaining quantities (– –) during the run time of the rotating circular vorticity profile.}
\label{fig:figure27}
\end{figure}

\paragraph*{}
First, we consider the VE fluid discussed in figure~\ref{fig:figure6}. The vorticies in the rotating vortex pair have equal
strengths $(\Gamma=5)$ and are separated by a distance of $d=3{\pi}$. Figure~\ref{fig:figure6} has been re-plotted in figure \ref{fig:figure25} with a circle of radius $4{\pi}$ units, which is displayed in the subplots. 
Figure \ref{fig:figure26}(a) represents the net change in $W$ due to transport terms: radiative $\bf{S}$, convective $\bf{T}$, and dissipative $\bf{P}$. We observe a continuous decay in the magnitude of $W$ rather than remaining constant. This suggests that the transport terms should change with time. The evolution of these time-varying terms is shown by the plots of the radiative term $\bf{S}$ in figure \ref{fig:figure26}(b), the convective term $\bf{T}$ in figure \ref{fig:figure26}(c), and the dissipative term $\bf{P}$ in figure \ref{fig:figure26}(d). It is clear from the vortex contour plot in figure \ref{fig:figure25} that the emerging waves and the rotating vortex pair remain inside the region for around the first t = 18 time units. In this time range (0-18), the zero contributions from the convective term $\bf{T}$ and the radiative term $\bf{S}$ also demonstrate this. Nevertheless, the dissipative term depicted in figure \ref{fig:figure26}(d) is primarily responsible for the constant decline in $W$ that happens throughout this time. However, as time goes on, after the first wavefront starts emerge after time 18, waves are continuously released that cross this circular boundary. Since the wave emission from one vortex considerably impedes the emission from the other, causing the somewhat continuous deviation in the center of rotation. This can be noticed from the fluctuating finite value of convection term $\bf{T}$.  Figure \ref{fig:figure26}(d) illustrates the role of the dissipating term $\bf{P}$, which almost becomes zero around 65. Thus, the comparative study of figures \ref{fig:figure26}(b), (c), and (d) shows that the TS waves swallow the vortex pair after around t = 65, which is also visible from the last panel of figure \ref{fig:figure25}.
In figure \ref{fig:figure27}, dWdt (solid line) and the total of all three terms $\bf{S+T+P}$ (dotted line) are plotted independently. As can be observed, the two curves precisely mirror one another, demonstrating that their sum equals zero as predicted by the equation (\ref{eq:integral_equ}).
\subsubsection{$\Gamma_1=5$, $\Gamma_2=5$; $a_1=0.5\pi$, $a_2=0.5\pi$; $and$ $d={2\pi}$}
\label{eq_strength5_eq_size_closely_STP}
\begin{figure}[!ht]
\centering 
\includegraphics[width=\textwidth]
{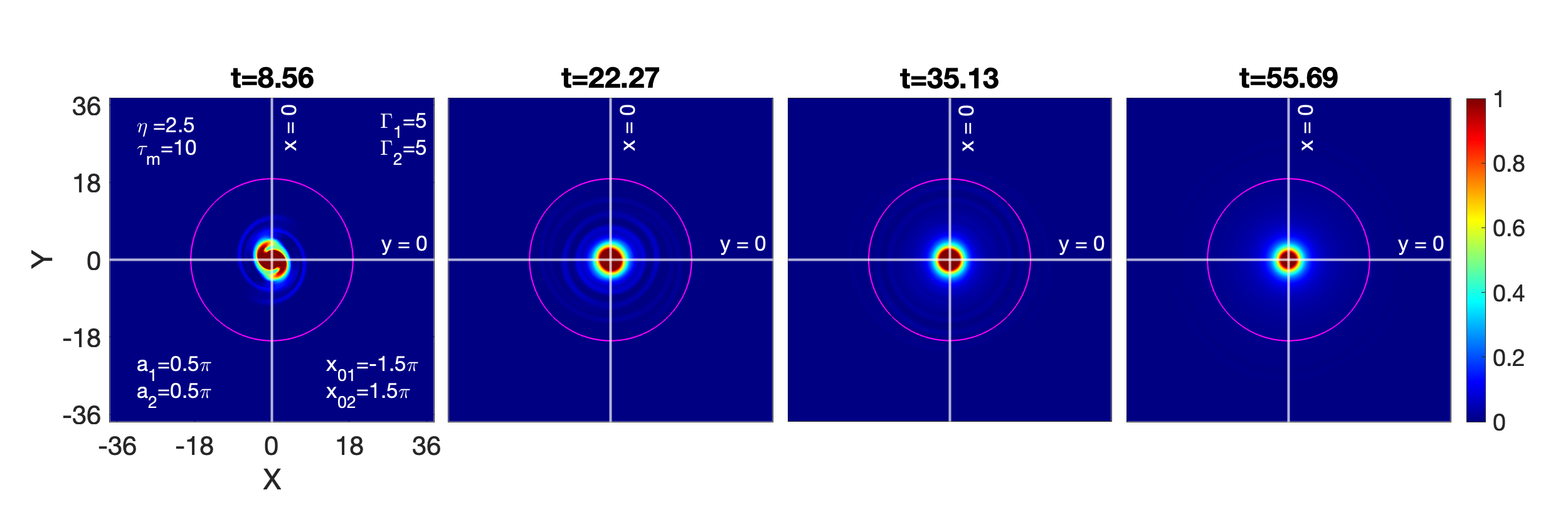}
\caption{Re-plot the figure~\ref{fig:figure10}(b). The time evolution of an identical vortex pair, equal circulation strength $\Gamma=5$, in viscoelastic fluid ($\eta=2.5$; $\tau_m=10$). A circular local volume element (inside the circumference) over which the different transport quantities are calculated.}
\label{fig:figure28}
\end{figure}

\paragraph*{}
We now take a look at the VE fluid that was covered in figure \ref{fig:figure10}(b). The revolving vortex pair's vorticies are spaced apart by $d=1.5{\pi}$ and have identical intensities $(\Gamma=5)$. The  figure \ref{fig:figure28} shows the re-plot of figure \ref{fig:figure10}(b) with a circle of radius $4{\pi}$ units.

\begin{figure}[!ht]
\centering 
\includegraphics[width=\textwidth]
{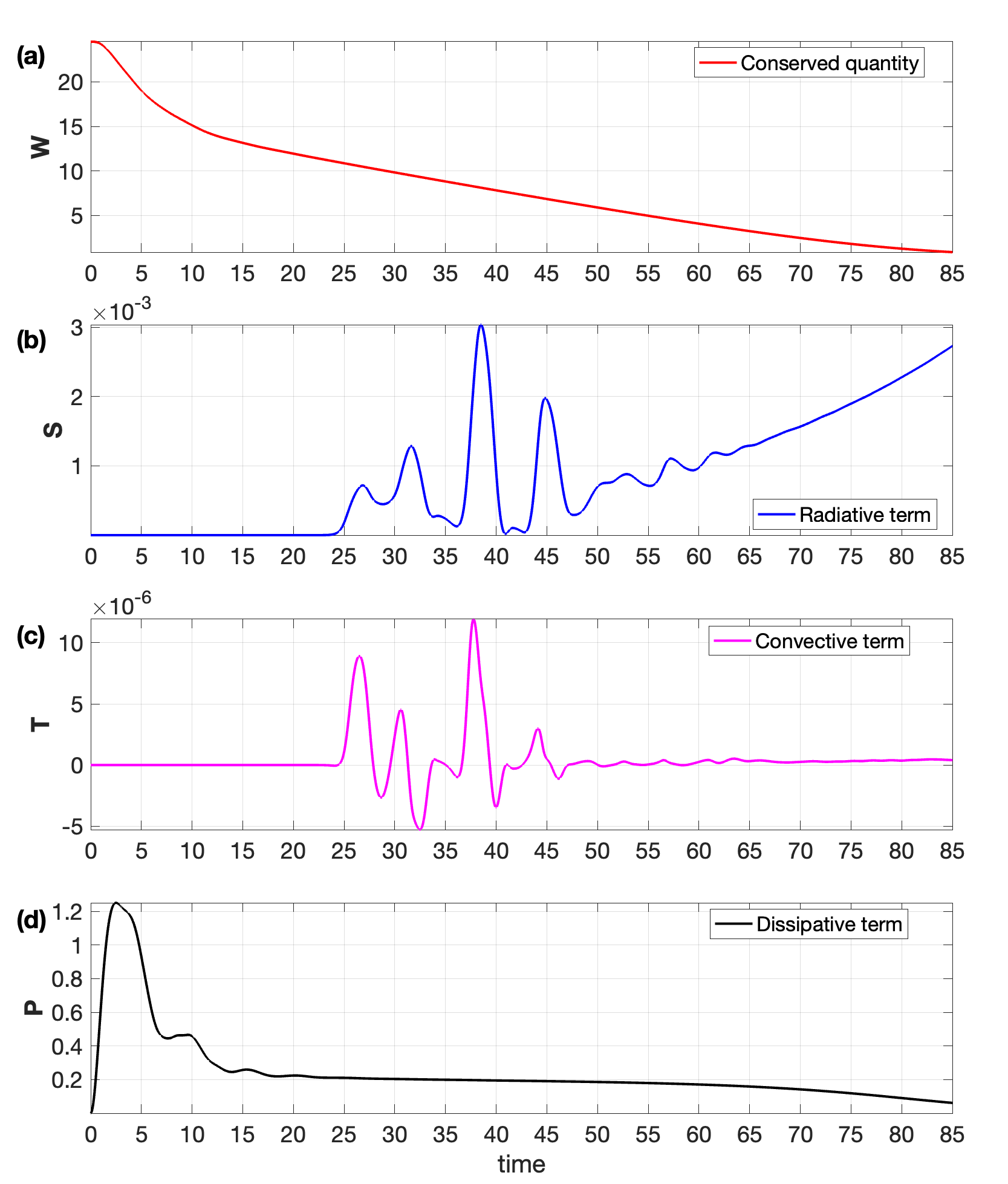}
\caption{The subplot (a) represents the change in W by wave emission. The contribution of convection is shown in subplot (b). The role of dissipating term is shown in subplot (c), which is observed to be finite.}
\label{fig:figure29}
\end{figure}

\begin{figure}[!ht]
\centering 
\includegraphics[width=\textwidth]
        {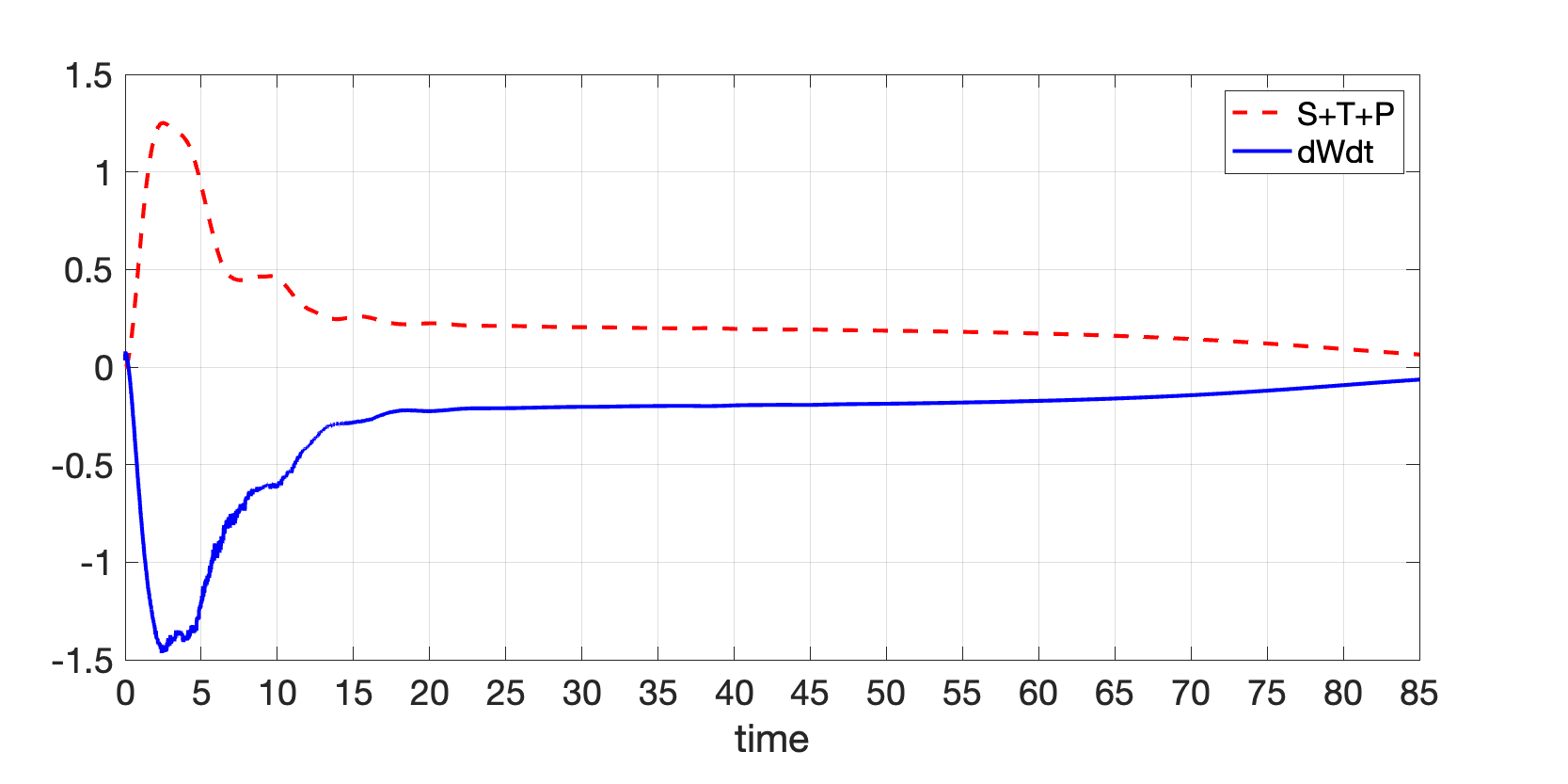}
\caption{Time derivative of conserved quantity W (–) is the mirror image to the total sum ($\bf{S+T+P}$) of all remaining quantities (– –) during the run time of the rotating circular vorticity profile.}
\label{fig:figure30}
\end{figure}

Everything occurs within of the circle for the about first $t = 25$ time units, as demonstrated by the vortex contour subplots in figure \ref{fig:figure28} and the zero contributions of $\bf{S}$ in figure \ref{fig:figure29}(b) and $\bf{T}$ in figure \ref{fig:figure29}(c). However, a consistent decline in $W$ transpires during this duration, mostly due to the dissipative component depicted in figure \ref{fig:figure29}(d).
In the time period from 25 to 50, the fluctuation in the value of the convection term $\bf{T}$ is due to the interplay between emitting waves and vorticies.  With time, the vortices gradually merge into a single symmetric vortex. This vortex stays stationary and and only revolves around its axis; there is no fluid convection across the region.
In Fig. \ref{fig:figure30}, the three terms' total (S + T + P) and dWdt (solid line) are plotted separately. The two curves exactly mirror each other, as can be seen, proving that their sum equals zero as the equation (\ref{eq:integral_equ}), which predicts this, indicates.
 Two more cases have been discussed in Appendix ~\ref{simulation_Conserved_quantity_A}.
\section{Conclusions and Outlook}
\label{Conclusions}
 \paragraph*{}
In this work, a well-recognized phenomenological generalized hydrodynamics fluid model has been used to describe the SCDPs as VE fluid media. Understanding the notion of merging through a co-rotating pair of vortices within the incompressible limit of VE fluid is the main objective of this essay. The transverse mode is the only mode that favors from the incompressible limit, while the acoustic mode is abstained. To examine the impact of transverse shear waves on merging phenomena, we take incompressible limit of the GHD model. The following are a few key findings:
\begin{enumerate}
\item {Vortex pair of equal strength and disparate size}
\vspace{-0.2cm}
\subitem {-~Unlike HD fluids, even for the widely positioned vortices, emerging shear waves facilitate the merging events in VE fluids.}
\vspace{-0.2cm}
\subitem {-~Specifically, widely positioned, the shear waves assist the merging in a medium that either has strong circulation vorticies or weak coupling strength, while in a medium that either has weak circulation vorticies or strong coupling strength, the shear waves make the vortex pair disappear prior to the merger happening.}
\vspace{-0.2cm}
\subitem {-~Closely positioned, the merger takes place for the medium with mild-strong coupling strength and medium with medium-strong coupling strength. On the other hand, in a medium with strong coupling strength, the structures disappear even before the merging
occurs.}
\vspace{-0.2cm}
\item {Vortex pair of disparate strength and equal size}
\subitem{ -~We find that the weaker vortex vanishes and the final vortex moves somewhat out of center for both the HD and VE fluids. This occurs more quickly in VE fluids than in HD fluids due to the presence of shear waves.}
\vspace{-0.2cm}
\subitem{ -~The disappearance of the weaker vortex in VE fluids is proportional to the coupling strength of the medium.}
\vspace{-0.2cm}
\item {Vortex pair of equal strength and disparate size}
\vspace{-0.2cm}
\subitem{ -~The process of merging for varying diameters is remarkably similar to the vortices of varying strengths,in which the smaller vortex acts like a weaker vortex and the larger vortex behaves like a strong vortex.}
\vspace{-0.2cm}
\item {Conserved equation}
\subitem{ -~it is numerically shown that the conservation law is satisfied with high accuracy where the conserved quantity W changes due to radiative, convective, and dissipative processes}
\end{enumerate}
 \paragraph*{}
We may state more broadly that by adjusting the coupling strength parameters of the medium, which usually represents the strength , one can control the evolution of the merging process and, consequently, control over transport features such as mixing and diffusion. In the present paper, we consider the homogeneous medium in terms of constant density, constant temperature, and constant charge. It would also be interesting to study the effect of heterogeneity (e.g., density, temperature, and charge)  on the merging steps. This study is left for future work. To validate this numerical work, an experimental research effort is needed. We think that our findings can motivate more research. It is noteworthy that the results obtained here are not restricted to the strongly coupled dusty plasma medium, but are generalizable to any visco-elastic medium.
\section{Acknowledgements}
This work was supported by the National Science Foundation through NSF-2308742 and NSF EPSCoR OIA-2148653.
\label{ack}
\appendix
 \section{Derivation of a conserved equation and quantity}
 \label{Conserved_quantity_A}
\paragraph*{}
In order to drive the conserved equation, the generalized momentum equation~(\ref{eq:momentum2}) has been formulated as a set of two coupled convective equations 
\begin{eqnarray}\label{eq:vort_incomp1_A}
 {\frac{\partial \vec{v}_d } {\partial t}+\vec{v}_d  \cdot{\nabla}\vec{v}_d } + \frac{\nabla {p_d}}{\rho_{cd}} -\frac{\rho_c}{\rho_{cd}} {\nabla {\phi_d}} ={\vec \psi}
\end{eqnarray}
\begin{equation}\label{eq:psi_incomp1_A}
	\frac{\partial {\vec \psi}} {\partial t}+\vec{v}_d \cdot \nabla{\vec \psi}=
	{\frac{\eta'}{\tau_m}}{\nabla^2}{\vec{v}_d }-{\frac{\vec \psi}{\tau_m}}{.}
\end{equation}
It is evident from Eq.~(\ref{eq:vort_incomp1_A}) that the strain produced in the elastic medium by the time-varying velocity fields is represented by the quantity ${\vec \psi}(x,y)$. Take the curl of equation~(\ref{eq:vort_incomp1_A}). The curls of second and third terms vanish because, because we have assumed a constant charge density, a gradient's curl is zero. Equation \ref{eq:vort_incomp1_A} is transformed into:
  \begin{eqnarray}\label{eq:vort_incomp3_A} 
\frac{\partial{\vec \omega}} {\partial t}+\vec{v}_d \cdot \nabla{{\vec \omega}}
={\nabla}{\times}{\vec \psi}{,}
\end{eqnarray}
\begin{eqnarray}\label{eq:psi_incomp3_A}
\frac{\partial {\vec \psi}} {\partial t}+\vec{v}_d \cdot \nabla{\vec \psi}=
{\frac{\eta'}{\tau_m}}{\nabla^2}{\vec{v}_d }-{\frac{\vec \psi}{\tau_m}}{.}
\end{eqnarray}
Later, for the numerical simulations, we will use the coupled equations (\ref{eq:vort_incomp3_A}) and (\ref{eq:psi_incomp3_A}). Now, take dot products (\ref{eq:vort_incomp3_A}) by ${\eta'}/{\tau_m}{\times}\vec{\omega}$ and (\ref{eq:psi_incomp3_A}) by $\vec{\psi}$, then add them
 \begin{eqnarray}
 \label{eq:gy_sum1_A}
 {\frac{\partial}{\partial{t}}}{\left(\frac{\psi^2}{2}+{\frac{\eta'}{\tau_m}}\frac{\omega_z^2}{2}\right)}+{\vec{\nabla}}{\cdot}{\frac{\eta'}{\tau_m}}({\omega_{z}}{{\times}{\vec \psi}})+ \nonumber \\
 {\nabla}{\cdot}{\vec{v}}\left({{\frac{\psi^2}{2}}}+{\frac{\eta}{\tau_m}}{\frac{\omega_z^2}{2}}\right)=-{\frac{\psi^2}{\tau_m}}
 \end{eqnarray}
The equation~(\ref{eq:gy_sum1_A}) looks Poynting like equation
 \begin{eqnarray}
 \label{eq:dgy_A}
 {\frac{\partial W}{\partial t}}+{\nabla}{\cdot}{\vec{S}}+{\nabla}{\cdot}{(T\vec {v})}+P_d =0
 \end{eqnarray}
Here, $W\equiv\left(\frac{\psi^2}{2}+{\frac{\eta'}{\tau_m}}\frac{\omega_z^2}{2}\right)$, 
${\vec{S}}\equiv{\frac{\eta'}{\tau_m}}({\omega_{z}}{{\times}{\vec \psi}})$, 
 $P_d\equiv{\frac{\psi^2}{\tau_m}}$, 
 and $T\equiv\left(\frac{\psi^2}{2}+{\frac{\eta'}{\tau_m}}\frac{\omega_z^2}{2}
 \right)$. 
 This shows that the rate of change of $W$ depends on dissipation through $P_d$ in the medium, a convective and radiative Poynting flux of $T \vec{v}$ and $\vec{S}$ respectively. Equation (\ref{eq:dgy_A}) can also be recast in the following integral form: 
 \begin{eqnarray}
 	{{{\frac{\partial}{\partial t}}}{\int_V}Wdv}+{\oint_{S}}{\vec S{\cdot}d{\vec{a}}}+{\oint_{S}}{T~\vec v{\cdot}d{\vec a}}=-{\int_V}{P_d}{dv}
 \end{eqnarray}
 which corresponds to 
\begin{multline}\label{eq:integral_equ_A} 
\underbrace{{\frac{\partial}{\partial t}}
{\int_{V}}{\left(\frac{\psi^2}{2}+{\frac{\eta}{\tau_m}}\frac{\omega_z^2}{2}
\right)}dv}_{\text{\bf {dWdt}}}= \\
-\underbrace{{\frac{\eta}{\tau_m}}{\oint_{S}}({\omega_{z}}{{\times}{ \vec
\psi}}){\cdot}d{\bf{a}}}_{\text{
\bf S}}-\underbrace{{\oint_{S}}\left(\frac{\psi^2 }{2 }+{\frac{
\eta}{\tau_m}}\frac{\omega_z^2}{2}\right){\vec{v}_d }{\cdot}d{\bf{a}}}_{\text{\bf
T}}-\underbrace{{\int_{V}}{\frac{\psi^2}{\tau_m }}dv}_{\text{\bf P}}{.} 
 \end{multline}
Let us analyse the physical meaning of each of the terms.  Here, ${\bf{dWdt}}$ is the sum of square integrals of vorticity ($\omega_z$) and the velocity strain ($\psi$, created in the elastic medium by the time varying velocity fields).  The radiation  term  ${\bf{S}}$ is like a Poynting flux for the radiation corresponding to the transverse shear waves. It accommodates the integral of the cross product of $\omega_z \hat{z}$ and $\vec{\psi}$. A comparison with electromagnetic light waves where $\vec{E} \times \vec{B}$ acts as a radiation flux shows that the corresponding  two fields here are $\omega_z \hat{z}$ and $\vec{\psi} $. The other mechanism causing the change in $W$ is through convection ${\bf{T}}$ and the viscous dissipation ${\bf{P}}$ through $\eta$. 
\paragraph*{}
Note: An earlier paper \cite{dharodi2016sub} has a complete, step-by-step derivation of this theorem as well as numerical simulations that validate it.
 \section{Continuation of the Section \ref{simulation_Conserved_quantity}: Numerical validation of the conserved  equation and quantity $W$}
\label{simulation_Conserved_quantity_A}
In this section, we continue the discussion from Section \ref{simulation_Conserved_quantity} and cover two additional instances.
\subsubsection{$\Gamma_1=5$, $\Gamma_2=3$; $a_1=0.5\pi$, $a_2=0.5\pi$; $and$~$d={3\pi}$}
\label{uneq_strength_g5g3_eq_size_widely_STP}
 \subsection{Numerical validation of the conserved  equation and quantity $W$}

\begin{figure}[!ht]
\centering 
\includegraphics[width=\textwidth]
{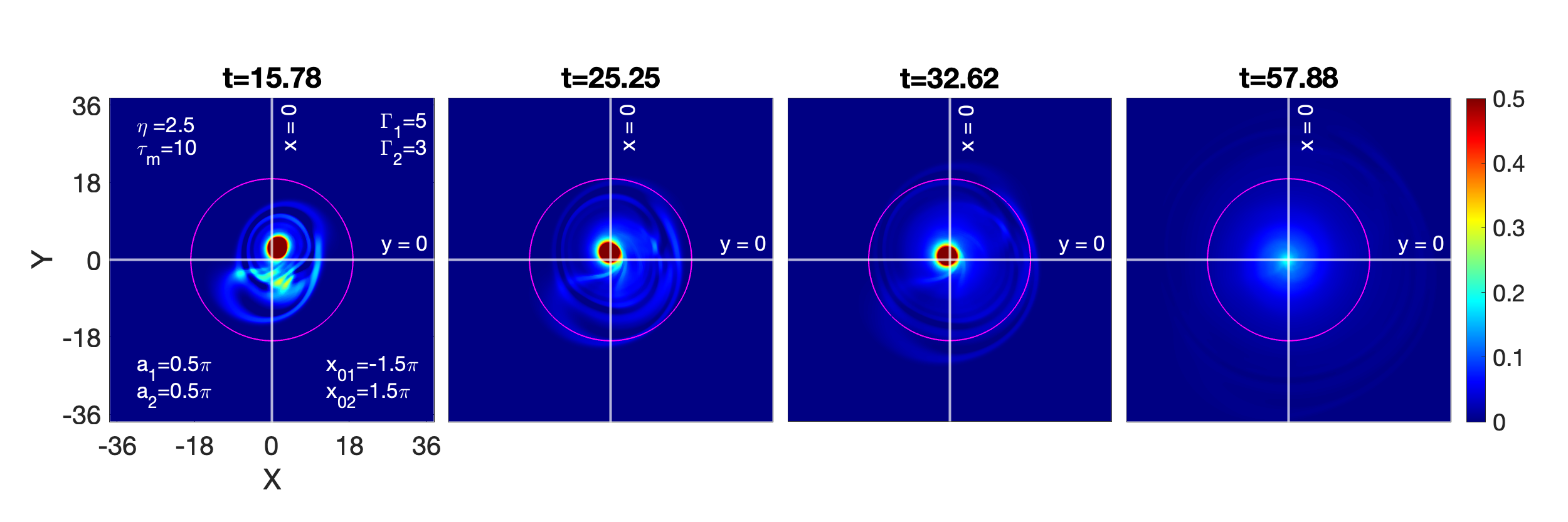}
\caption{Re-plot the figure~\ref{fig:figure19}(b). The time evolution of an identical vortex pair, equal circulation strength $\Gamma=5$, in viscoelastic fluid ($\eta=2.5$; $\tau_m=10$). A circular local volume element (inside the circumference) over which the different transport quantities are calculated.}
\label{fig:figure31}
\end{figure}

\begin{figure}[!ht]
\centering 
\includegraphics[width=\textwidth]
{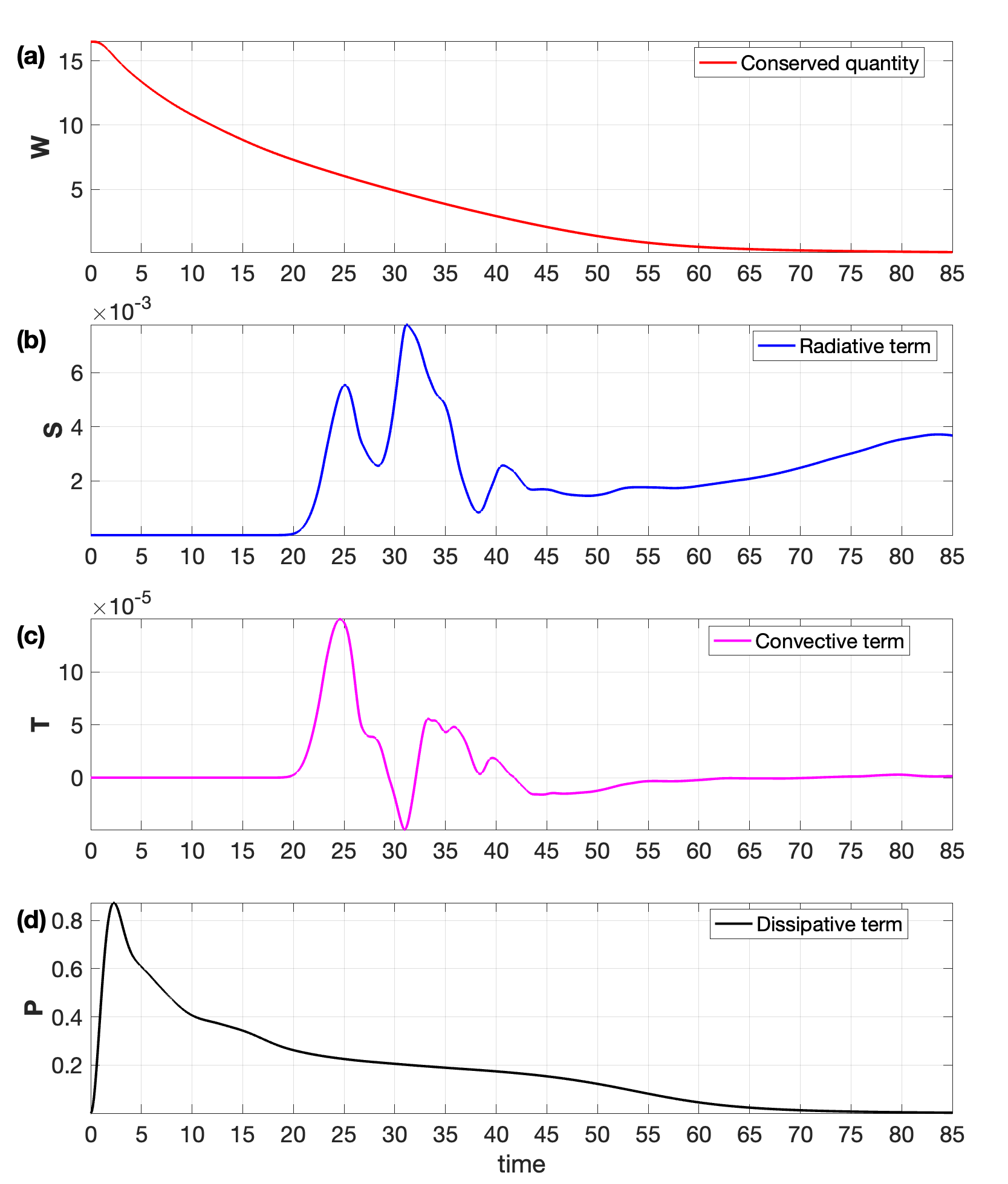}
\caption{The subplot (a) represents the change in W by wave emission. The contribution of convection is shown in subplot (b). The role of dissipating term is shown in subplot (c), which is observed to be finite.}
\label{fig:figur32}
\end{figure}

\begin{figure}[!ht]
\centering 
\includegraphics[width=\textwidth]
        {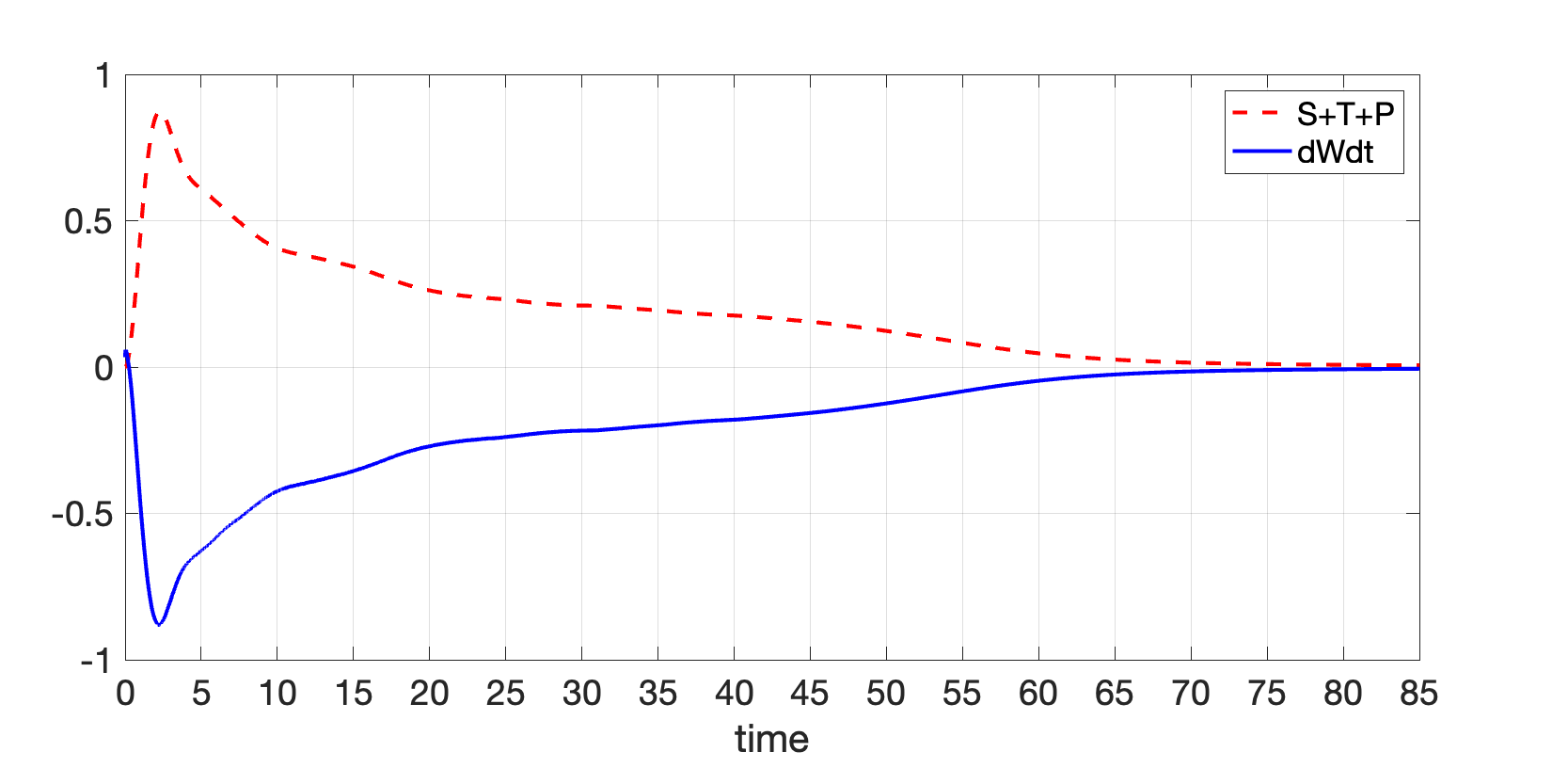}
\caption{Time derivative of conserved quantity W (–) is the mirror image to the total sum ($\bf{S+T+P}$) of all remaining quantities (– –) during the run time of the rotating circular vorticity profile.}
\label{fig:figure33}
\end{figure}

Now let's examine the VE fluid depicted in the figure \ref{fig:figure19}(b). The two vorticies have disparate strengths of $\Gamma_1=5$ (stronger) and $\Gamma_2=3$ (weaker). The inial distance between these vorticies is $d=3{\pi}$. The re-plot of figure \ref{fig:figure19}(b) with a circle of radius $4{\pi}$ units is depicted in figure \ref{fig:figure31}.
Initially, as in the earlier situations, the only dissipative term lowered the magnitude of $W$ until the interaction took place in the circular boundary. In the time period from 20 to 55, continuous emission and fluid convection across the region are observed. Eventually, the vortex pair will be engulfed by the emerging shear waves. It can be observed through contour plot visualization in figure \ref{fig:figure31} and the analysis of different transport terms in figure \ref{fig:figur32}.
In Fig. \ref{fig:figure33}, dWdt (solid line) and the sum of the three terms S + T + P (dotted line) are plotted separately. As can be seen, the two curves exactly mirror one another, proving that the equation's prediction that their sum equals zero is correct (\ref{eq:integral_equ}).
\subsubsection{$\Gamma_1=5$, $\Gamma_2=5$; $a_1=0.6\pi$, $a_2=0.4\pi$; $and$~$d={3\pi}$}
\label{eq_strength5_uneq_size_widely_STP}
\paragraph*{}
Finally, we consider the case of two vorticies with disparate sizes of $a_1=0.6{\pi}$ (larger) and $a_2=0.4{\pi}$ (smaller) of equal strength $\Gamma_1{=}\Gamma_2=5$ and separated by distance $b_0=6a_0$, discussed in figure \ref{fig:figure24}(b). The re-plot of this figure with a circle of radius $4{\pi}$ units is shown in figure \ref{fig:figure31}.
\begin{figure}[!ht]
\centering 
\includegraphics[width=\textwidth]
    {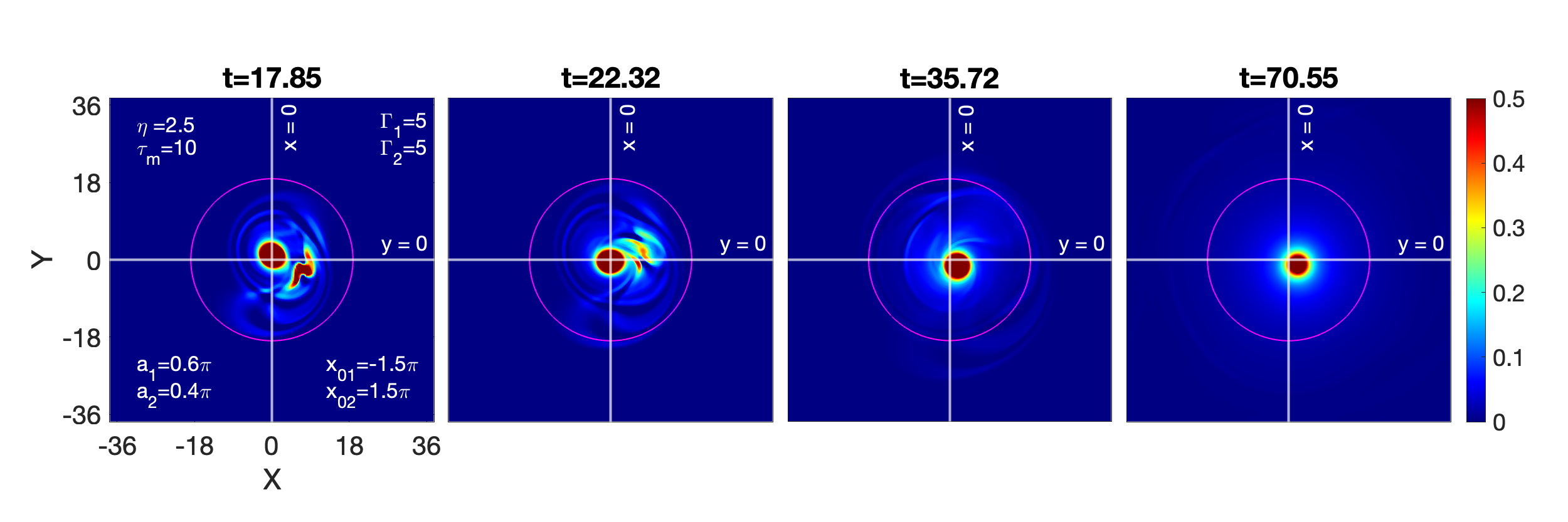}
       \caption{Re-plot the figure~\ref{fig:figure24}(b). The time evolution of an identical vortex pair, equal circulation strength $\Gamma=5$, in viscoelastic fluid ($\eta=2.5$; $\tau_m=10$). A circular local volume element (inside the circumference) over which the different transport quantities are calculated.}
\label{fig:figure34}
\end{figure}
As in the earlier situations, initially, the the magnitude of $W$ get lowered due to the only dissipative term. In intermediate time  period from 20 to 55, continuous wave emission and fluid convection across the region are observed. Eventually, the vortex pair merge into a single symmetric vortex which  continuously radiate the share waves. It can be visualized through contour subplots in figure \ref{fig:figure31} and by the analysis of different transport terms in figure \ref{fig:figure35}.
\begin{figure}[!ht]
\centering 
\includegraphics[width=\textwidth]
        {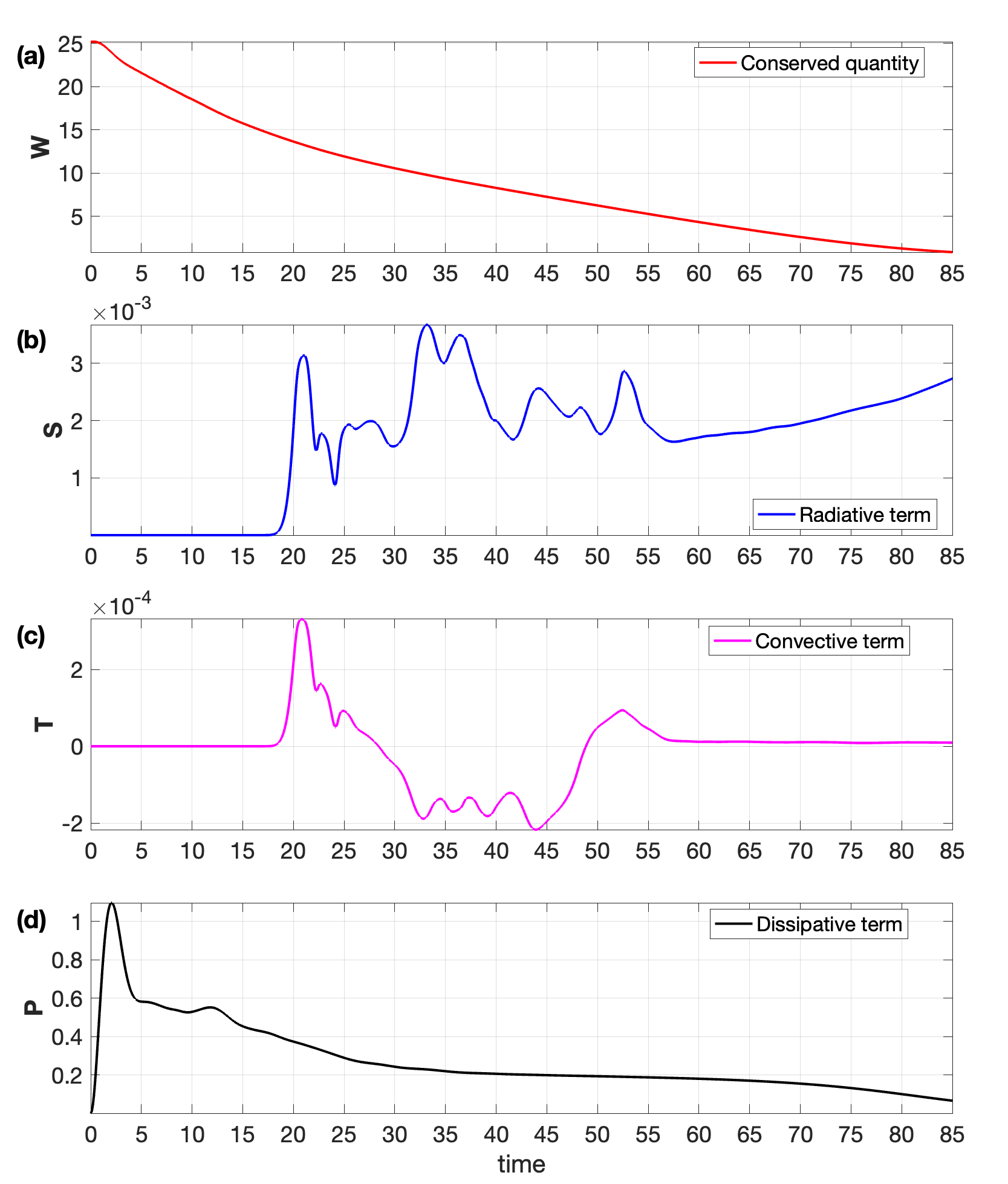}
        \caption{The subplot (a) represents the change in W by wave emission. The contribution of convection is shown in subplot (b). The role of dissipating term is shown in subplot (c), which is observed to be finite.}
  \label{fig:figure35}
\end{figure}
The total of the three terms (S + T + P) and dWdt (solid line) are presented individually in Fig. \ref{fig:figure36}. As can be observed, the two curves perfectly mirror one another, demonstrating that their sum equals zero as shown by equation (\ref{eq:integral_equ}), which makes this prediction.
\begin{figure}[!ht]
\centering 
\includegraphics[width=\textwidth]
    {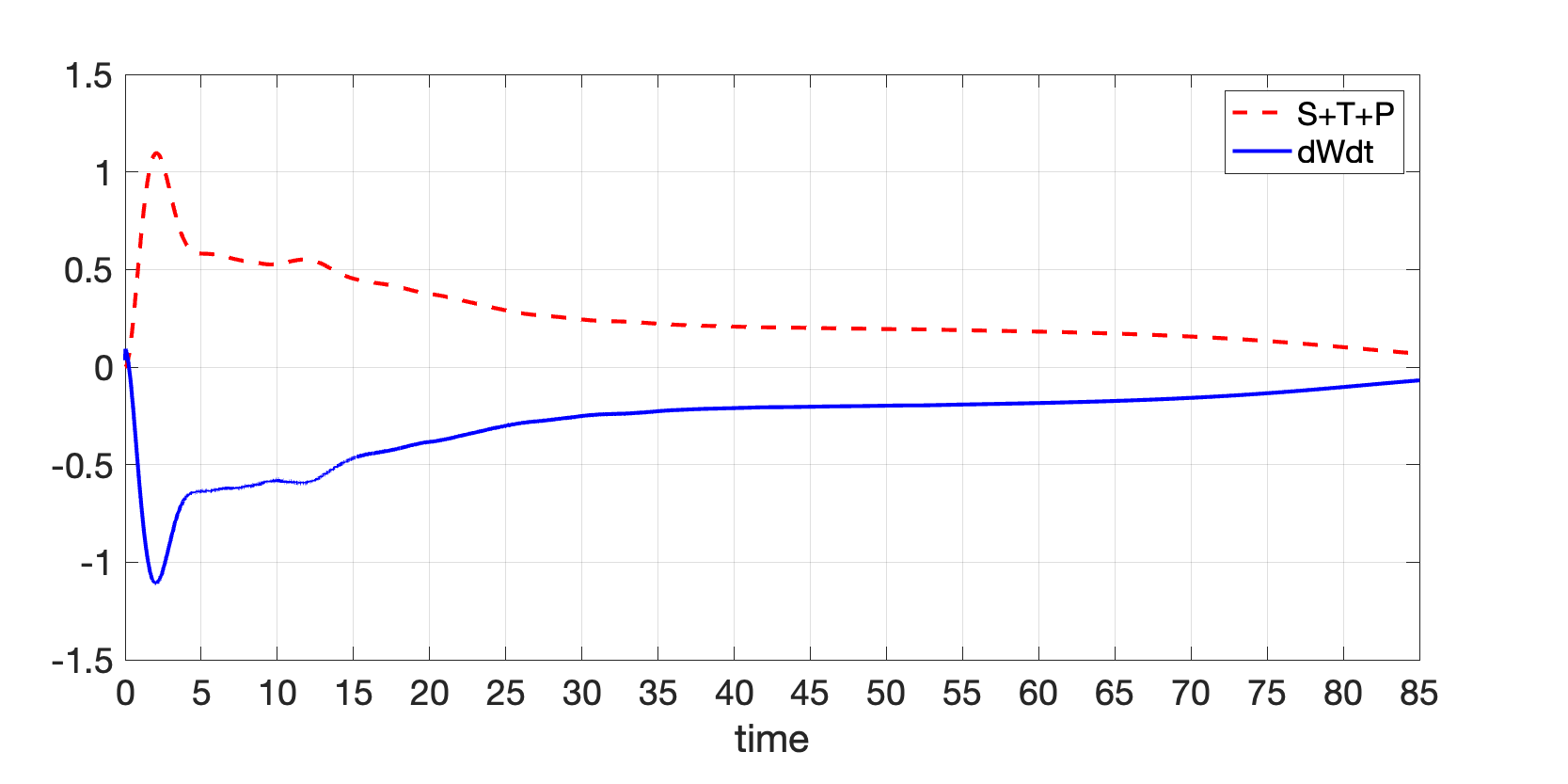}
\caption{Time derivative of conserved quantity W (–) is the mirror image to the total sum ($\bf{S+T+P}$) of all remaining quantities (– –) during the run time of the rotating circular vorticity profile.}
 \label{fig:figure36}
\end{figure}
\FloatBarrier


\end{document}